\documentclass[11pt]{article}

\usepackage{arxiv}

\usepackage[utf8]{inputenc} 
\usepackage[T1]{fontenc}    
\usepackage{hyperref}       
\usepackage{amsmath}        
\usepackage{lipsum}
\usepackage{subcaption}
\usepackage{graphicx}
\usepackage{natbib}
\usepackage{booktabs}
\usepackage{enumitem}
\usepackage{algorithm}
\usepackage[noend]{algpseudocode}
\usepackage{newtxtext,newtxmath}
\usepackage{bm}

\title{Simplified Vehicle-Bridge Interaction for Medium to Long-span Bridges Subject to Random Traffic Load}

\author{
  Soheil S. Eshkevari\thanks{ses516@lehigh.edu} \\
  Department of Civil and Environmental Engineering\\
  Lehigh University\\
  Bethlehem, PA 18015 \\
  \texttt{ses516@lehigh.edu} \\
  \And
  Thomas J. Matarazzo\thanks{tomjmat@mit.edu} \\
  Senseable City Lab\\
  Massachusetts Institute of Technology\\
  Cambridge, MA 02139 \\
  Information Science\\
  Cornell Tech\\
  New York, NY 10044\\
  \texttt{tomjmat@mit.edu} \\
  \And
 Shamim N. Pakzad \\
  Department of Civil and Environmental Engineering\\
  Lehigh University\\
  Bethlehem, PA 18015 \\
  \texttt{pakzad@lehigh.edu} \\
  }
  
\begin{document}


\maketitle

\begin{abstract}
This study introduces a simplified model for bridge-vehicle interaction for medium- to long-span bridges subject to random traffic loads. Previous studies have focused on calculating the exact response of the vehicle or the bridge based on an interaction force derived from the compatibility between two systems. This process requires multiple iterations per time step per vehicle until the compatibility is reached. When a network of vehicles is considered, the compatibility equation turns to a system of coupled equations which dramatically increases the complexity of the convergence process. In this study, we simplify the problem into two sub-problems that are decoupled: (a) a bridge subject to a random Gaussian excitation, and (b) individual sensing agents that are subject to linear superposition of the bridge response and the road profile roughness. The study provides sufficient evidences to confirm the simulation approach is valid with minimal error when the bridge span is medium to long, and the spatio-temporal load pattern can be modeled as random Gaussian. Quantitatively, the proposed approach is over 1,000 times more computationally efficient when compared to the conventional approach for a 500 m long bridge, with response prediction errors below $0.1\%$. 
\end{abstract}

\section{Introduction}

The problem of vehicle-bridge interaction (VBI) has been studied widely over recent years due to the broad applications spanning from fatigue analysis and bridge mobile sensing \citep{chen2007equivalent,zhu2015structural,zhu2016recent,yang2018state,eshkevari2020deconv} to ride comfort and safety analysis \citep{zhou2016vehicle,camara2019complete}. The complexity of the problem has resulted in a reliance on numerical modeling to evaluate research hypotheses \citep{yang2004vehicle,malekjafarian2014ident,sadeghi2019modal}. Consequently, today various numerical tools for VBI modeling are available, yet the majority are geared towards problems concerning individual vehicle dynamics, e.g., a single vehicle's interaction with a simple bridge. Recent applications on vehicle fleets and crowdsensing methods \citep{o2019quantifying, matarazzo2018crowdsensing} have provided insight into the wealth of SHM information that can be produced by ubiquitous mobile sensors. Such large-scale analyses call for interaction methods that can incorporate vehicular networks, everyday traffic scenarios, and are computationally efficient.

\subsection{Crowdsensing the Built Environment with Mobile Sensors}

\begin{figure}[!ht]
\centering\includegraphics[width=0.8\linewidth]{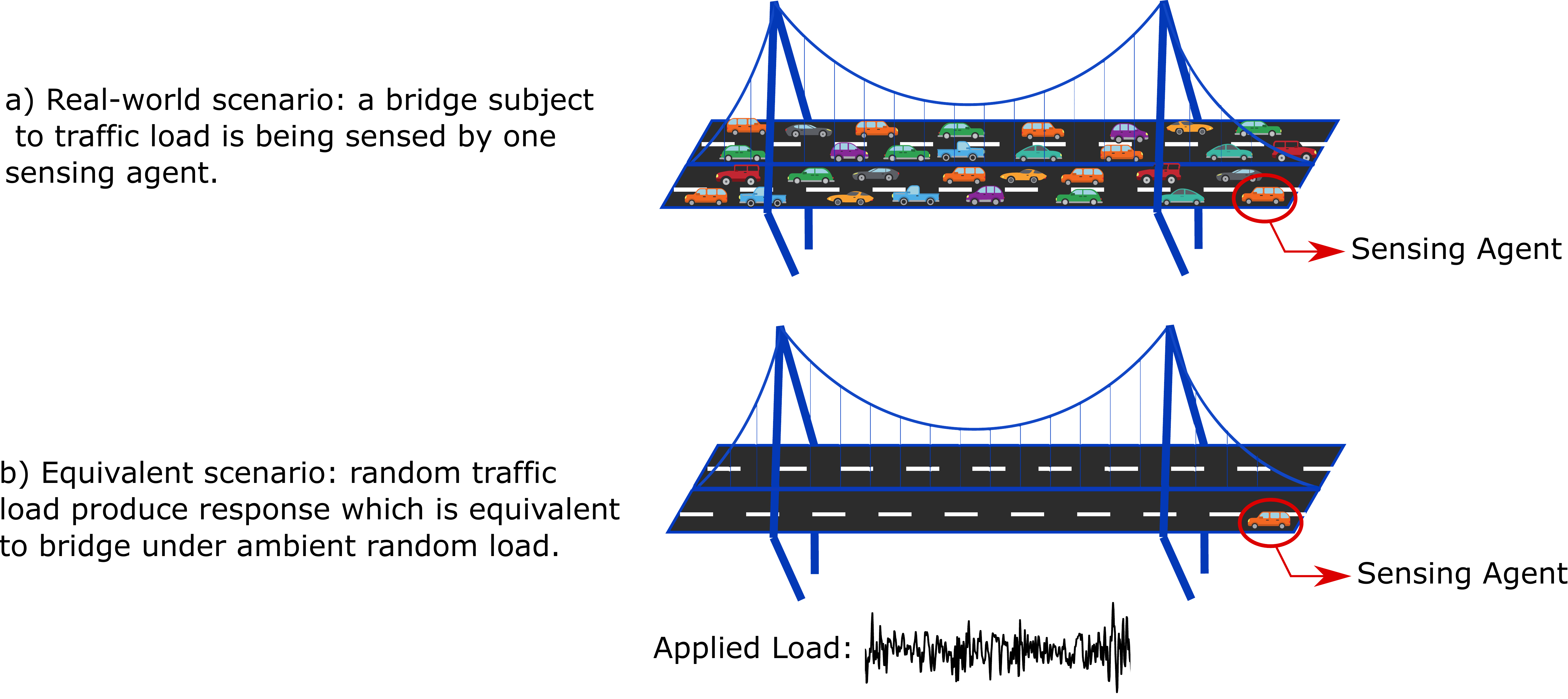}
\caption{Crowdsourcing framework. The sensing agent is one (or more) particular vehicle within a large pool of crossing vehicles. The problem is equivalent to a case in which the bridge is subject to ambient random load while being scanned by the sensing agent.}
\label{fig:problemOfInterest}
\end{figure}

The growing adoption of \textit{internet of things} technologies and connected devices in smart cities suggest a new sensing paradigm in which new information is regularly gathered from the crowd, e.g., individual smartphones, vehicular sensor networks, etc. \cite{calabrese2010real}  proposed a real-time data aggregation solution for constructing a dynamic urban map of large cities using crowdsourced smartphone data. \cite{wang2012understanding} quantified traffic patterns and proposed management applications based on large-scale mobile phone data. \cite{yu2015initial} successfully utilized smartphone sensors for structural health monitoring application due to its availability and inexpensive data acquisition. \cite{feng2015citizen,ozer2015citizen} also suggested novel applications in post-event bridge vibration analysis using stationary smartphones as sensors. Figure \ref{fig:problemOfInterest} illustrates how a vehicle can act as a sensing agent amongst bridge traffic. 
  \par
 
Crowdsensing inherently relies on \textit{mobile sensor networks}, which is an emerging data acquisition technique in structural health monitoring (SHM). Historically, observations of structural dynamics have been based on measurements collected by \textit{fixed sensor networks}. \cite{matarazzo2016stride,matarazzo2016structural} presented the STRIDE modal identification algorithm and verified that mobile sensor data was suitable for a comprehensive modal identification (frequencies, damping ratios, and mode shapes). \cite{matarazzo2016truncated} proposed the truncated physical state-space model as an efficient approach for representing time-space observations from a mobile sensor network. \cite{matarazzo2018scalable} presented an identification algorithm called STRIDEX to identify truncated physical model parameters, which enabled efficient and scalable modal identification using mobile sensors; it was shown that one mobile sensor provided a mode shape density comparable to 120 fixed sensors. 

\cite{matarazzo2018crowdsensing} presented a real-world application of mobile sensors, in the form of smartphones in moving vehicles. Significant indicators of the first three modal frequencies of the Harvard Bridge were found by aggregating data from about forty bridge trips. \cite{eshkevari2020signal,eshkevari2020mimc} proposed a method called MIMC to consider vibration data collected by multiple mobile sensors with random motions which successfully identified bridge modal properties in simulated applications.  These studies show promise for the use of crowdsourcing in bridge health monitoring. Yet further development is needed, e.g., analytical and experimental studies, to attain the sophistication and robustness of the traditional modal identification methods based on fixed sensor data. \par
 
\subsection{Vehicle-bridge interaction modeling}

More practical approaches for bridge health monitoring such as crowdsensing requires a computationally scalable numerical framework. A comprehensive literature review of common VBI simulation approaches is provided by \cite{gonzalez2010vehicle}. Initially, the vehicle-bridge interaction was modeled using 1D continuous beam models subject to simple moving loads \citep{fryba2013vibration} which is solvable in closed-form. By further development of computers and increasing use of the finite element method, the problem was reframed as a multi degrees of freedom (MDOF) system for the bridge interacting with simplified dynamical models of the vehicle. This approach has been broadly adopted for VBI modeling, mostly for short to mid-span bridges subject to a very limited number of vehicles with controlled motions. In this approach, once the models for the vehicle and the bridge are selected (based on required accuracy and fidelity), the dynamic equations of each component are separately built, in which the interaction forces between the vehicle and the bridge are coupled to the both sets of equations. Therefore, a numerical solver is required to solve the problem either iteratively or as a coupled system of equations. \par

The underlying principles of the approach, that is the interactive dynamic force acting between the vehicle and the bridge, have remained consistent throughout the literature. The uncoupled iterative algorithm is the most common method for VBI problems \citep{lin2005use,kim2008pseudo,obrien2010characteristic,gonzalez2012identification,yang2018state}. Various versions of the algorithm has been developed based on the problem requirements, e.g., different vehicle models, single DOF, quarter-car, or half-car models as well as different bridge models with different fidelity levels (such as 2D, 3D, with or without material or geometrical nonlinearities). However, in the majority of these studies, a short- to mid-span bridge has been considered. As mentioned in \cite{gonzalez2010vehicle}, when the vehicle mass is negligible compared to the bridge mass (which is the case for medium to long bridges) and a smooth pavement is assumed, a moving mass model can be replaced with the dynamic model of the vehicle that simplifies the simulation process. Road irregularities contribute complex dynamics to the interaction force, which emphasize the importance of a fully coupled model. \par

In the uncoupled iterative approach, the bridge model is analyzed multiple times (once at the beginning, and at least once for each time step inside the compatibility convergence loop). In addition, as the bridge dimension grows, an accurate bridge model requires more degrees of freedom, which increases the computational costs. A limited number of studies have considered long-span bridges along with a dense vehicle network for the simulation purpose. \cite{camara2019complete} recently modeled wind-bridge-vehicle interaction using the uncoupled iterative approach. The study could accurately model the system by adopting complex models for each component. The complexity of the approach implies that it requires great efforts to built such a high fidelity model, which may neither be a feasible nor cost effective solution for crowdsensing or other crude vehicle-bridge interacting scenarios. Moreover, bridge standards recommend lower dynamic factors for medium to long bridges compared to short bridges \citep{aashto2008bridge}; which means that the VBI interaction force is less dynamic and more similar to a constant moving load. These challenges and specifications suggest that it may not be required to use rigorous iterative solutions for VBI simulation of medium to long bridges subject to heavy traffic loads. This study intends to demonstrate that a simplified simulation approach inspired by the conventional uncoupled iterative algorithm \citep{gonzalez2010vehicle} is able to simulate VBI problems with high accuracy and dramatically less computational effort.

Figure \ref{fig:problemOfInterest} shows how the same notion is applicable in the VBI simulation. This figure demonstrates a scenario of interest in which the bridge is subject to a random traffic network. The objective is to simulate the system and finally calculate the collected response of the sensing agent. In a brute-force approach, the spatial coordinates and mechanical properties of every single vehicle in the network is required to fully determine the complex model. Such an accurate information setting is quite impractical and unnecessary. Alternatively, one can simulate the collective loading effect of the vehicle network (the sensing agent excluded) by ambient random load (as shown in Figure \ref{fig:problemOfInterest} - b). If the spatio-temporal ambient random load is represented as a matrix $\bm{F}_0$, the conventional algorithm for simulating the VBI problem is as shown in Algorithm \ref{alg:conventional}.

\begin{algorithm}[!ht]
	\caption{Conventional iterative VBI simulation.}
	\label{alg:conventional}
	\small \small
	\begin{algorithmic}[1]
		\State \textbf{Input}: $\bm{M}_{brg},\bm{C}_{brg},\bm{K}_{brg},\bm{M}_{vcl},\bm{C}_{vcl},\bm{K}_{vcl},\bm{F}_0,rgh$
		\State $\bm{Y}_{brg} = Newmark\beta(\bm{M}_{brg},\bm{C}_{brg},\bm{K}_{brg},\bm{F}_0)$	\For {t = 1, $\cdots$, T}
		\State
		Initiate $r:=1, r_n:=1000$
		\While {$abs(r-r_n) < threshold$} 
    	\State $r = \bm{Y}_{brg}(t)$
    	\State $wv = rgh(t) + r$
    	\State $wv^\prime = rgh^\prime(t) + \bm{Y}_{brg}^\prime(t)$
    	\State $y_{vcl}(t) = ODE45(\bm{M}_{vcl},\bm{C}_{vcl},\bm{K}_{vcl},wv,wv^\prime)$
    	\State $F_t = -\bm{K}_{vcl}(2)*(y_{vcl}(t)-wv) - \bm{C}_{vcl}(2)*(y_{vcl}^\prime(t)-wv^\prime)$
    	\State $R = - \bm{M}_{vcl}g - F_t$
    	\State $\bm{F} = \bm{F}_0$
    	\State $\bm{F}(t) = R$
    	\State $\bm{Y}_{brg} = Newmark\beta(\bm{M}_{brg},\bm{C}_{brg},\bm{K}_{brg},\bm{F})$
        \State $r_n = \bm{Y}_{brg}(t)$		
		\EndWhile
		\EndFor
        \State \textbf{Return} $\bm{Y}_{brg}, y_{vcl}$
	\end{algorithmic}
\end{algorithm}	

In this algorithm, $\bm{M}_{brg},\bm{C}_{brg},\bm{K}_{brg}$ and $\bm{M}_{vcl},\bm{C}_{vcl},\bm{K}_{vcl}$ characterize mechanical properties of the bridge and the vehicle, respectively. $rgh$ is a vector of roughness profile elevations at bridge DOFs. The algorithm performs the following steps:

\begin{enumerate}
    \itemsep0em 
    \item The bridge is subjected to random ambient load $\bm{F}_0$ at different physical locations.
    \item A vehicle starts moving from one side of the bridge and at each time instance, the bridge response (displacement) from the previous step in addition to the local roughness intensity (i.e., $rgh(t)$) is input to the vehicle system.
    \item The vehicle response to the applied force from the previous step is then analyzed using a Matlab ordinary differential equation (ODE) solver to calculate its displacement response (line 9 in Algorithm \ref{alg:conventional}). Based on this response, the interacting force between the sensing vehicle and the bridge is calculated as: $F_t = -\bm{K}_{vcl}[2](y_{vcl}(t)-wv) - \bm{C}_{vcl}[2](y_{vcl}^\prime(t)-wv^\prime)$ (where $[2]$ stands for the $2nd$ DOF of the vehicle, i.e., the tire). Note that if $F_t<0$, it is replaced with zero since it means that the vehicle lost its contact. 
    \item The interaction force from the vehicle to the bridge $F_t$ upgrades the original loading matrix $\bm{F}_0$ to produce $\bm{F}$. At this location, the bridge is required to be analyzed again with the updated force matrix. Here, Newmark-$\beta$ method is used for bridge dynamics analysis \citep{newmark1959method}. 
    \item If the difference between the updated bridge displacement and the one that was applied in Step 2 is higher than a predefined threshold, the process should be repeated from Step 2 onward by the updated bridge response. Otherwise, the vehicle moves to the next DOF on the bridge.
\end{enumerate}

Step 5 in this process (i.e., the \texttt{while} loop in Algorithm \ref{alg:conventional}) is expensive since it results in multiple full bridge analysis iterations within a time step. This is quite significant when the bridge is discretized with a large number of DOFs. Figure \ref{fig:simulTypes} summarizes the approaches one can take for calculation of the sensing vehicle's measurement. In case (a), the brute-force approach is shown in which all the vehicles are coupled with the bridge.

\subsection{Simplified Model}

This study proposes a fast and accurate simulation approach for VBI problems in which: (1) the bridge span is medium  to long and it is flexible, and (2) the vehicle network load is modeled as a random  spatio-temporal load over the bridge span. The second condition refers to the ambient vibrations caused by a network of moving vehicles \citep{de2000benchmark,ren2004roebling,ren2004output,pakzad2008design}. \par

\begin{figure}[!h]
\centering\includegraphics[width=0.8\linewidth]{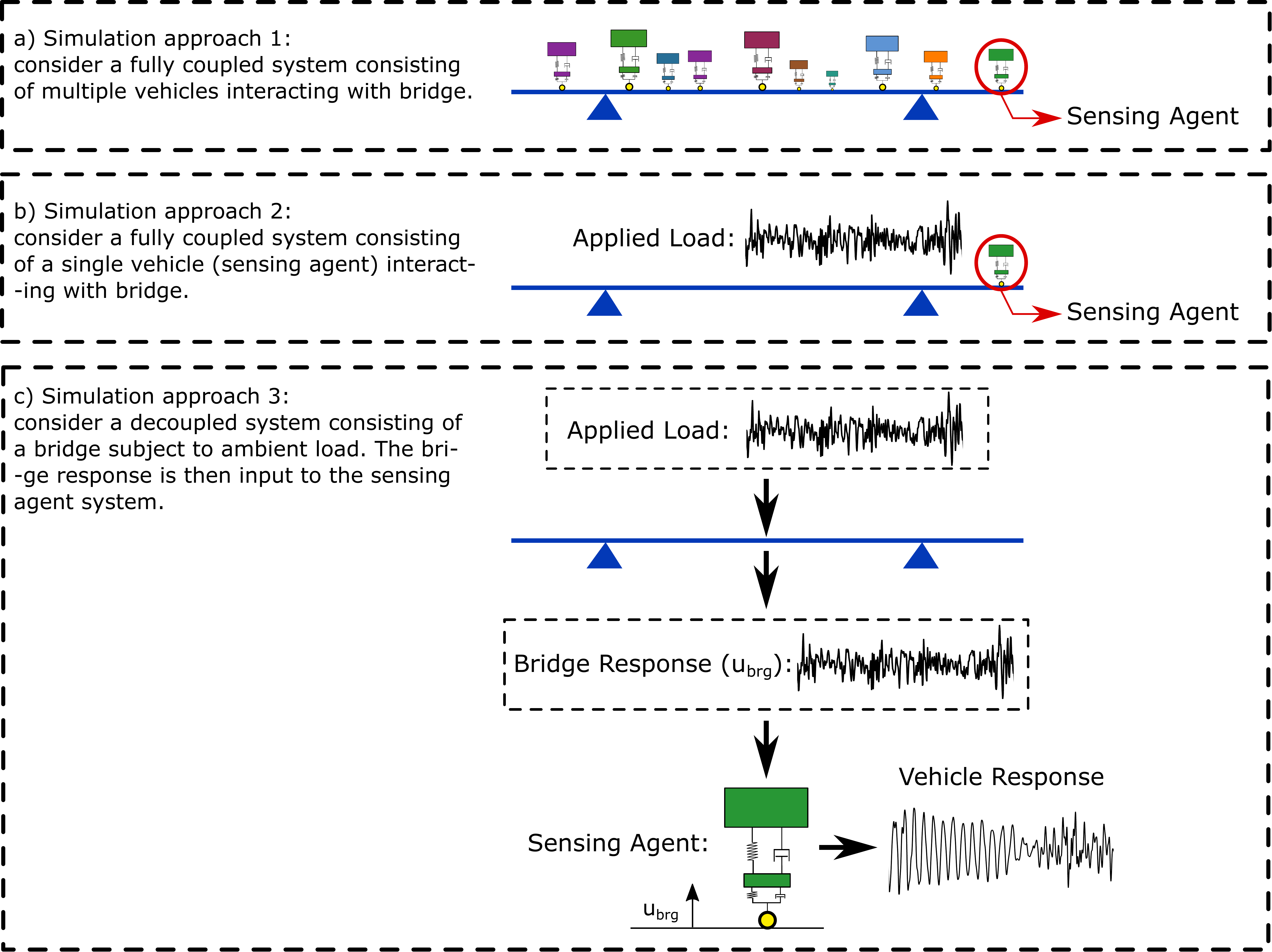}
\caption{Simulation approaches: a) a complex and coupled system of a vehicle network interacting with a bridge; b) a coupled system of the sensing vehicle interacting with the bridge. The bridge is separately subject to an ambient load to capture the vehicle network's load; c) the proposed approach in which the bridge is only subject to the ambient load. The response is then applied to an uncoupled model of the sensing vehicle to produce the vehicle output.}
\label{fig:simulTypes}
\end{figure}

 In Figure \ref{fig:simulTypes}b, shows a simplified representation of Figure \ref{fig:problemOfInterest}a, in which the traffic network (the sensing agent excluded) is replaced with an applied ambient white noise load while the sensing agent is still interacting with the bridge in a coupled fashion. While this approach is significantly less computationally expensive by comparison, the coupled system still requires iterations to reach the compatibility between the vehicle and the bridge at each time step. In this paper, we present an approach in which the compatibility calculations between two interacting components are not iterative, as shown in Figure \ref{fig:simulTypes} - c. In this approach, we posit that the dynamical effect of an individual sensing agent on a bridge response is negligible when the bridge is medium to long and the cumulative effect of other loads (the individual vehicle excluded) is significantly greater than a single vehicle. The approach is presented in Algorithm \ref{alg:simplified}:

\begin{algorithm}[!h]
		\caption{Simplified non-iterative VBI simulation.}
		\label{alg:simplified}
		\small \small
		\begin{algorithmic}[1]
			\State \textbf{Input}: $\bm{M}_{brg},\bm{C}_{brg},\bm{K}_{brg},\bm{M}_{vcl},\bm{C}_{vcl},\bm{K}_{vcl},\bm{F}_0,rgh$
			\State $\bm{Y}_{brg} = Newmark\beta(\bm{M}_{brg},\bm{C}_{brg},\bm{K}_{brg},\bm{F}_0)$
			\For {t = 1, $\cdots$, T}
        	\State $r = \bm{Y}_{brg}(t)$
        	\State $wv = rgh(t) + r$
        	\State $wv^\prime = rgh^\prime(t) + \bm{Y}_{brg}^\prime(t)$
        	\State $y_{vcl}(t) = ODE45(\bm{M}_{vcl},\bm{C}_{vcl},\bm{K}_{vcl},wv,wv^\prime)$
			\EndFor
            \State \textbf{Return} $\bm{Y}_{brg}, y_{vcl}$
		\end{algorithmic}
	\end{algorithm}	 
	
In this algorithm, the bridge is only analyzed once at the beginning under $\bm{F}_0$. The bridge response is then linearly superimposed with $rgh$ and then, applied to the vehicle dynamical model. In fact, the approach is similar to the constant force method proposed in \cite{gonzalez2010vehicle}. However, in our approach the vehicle dynamics is incorporated in the vehicle response, which was not the case in a moving mass model. The approach has not been proposed or utilized previously; yet needs to be fully justified and evaluated. In the rest of this paper, we first propose a theoretical proof on a simplified case of the coupled VBI problem. This part intends to demonstrate that bridge to vehicle mass and stiffness ratios are the keys to determine the coupling degree. In the next step, VBI responses of multiple bridges with different characteristics and vehicles are numerically simulated using coupled (i.e., conventional) and uncoupled (i.e., simplified) procedures and results are compared. Discussions and comparison of the numerical results are also supplemented in the last sections. 

\section{Theoretical Approach}\label{sec:theory}

\begin{figure}[!ht]
\centering\includegraphics[width=0.6\linewidth]{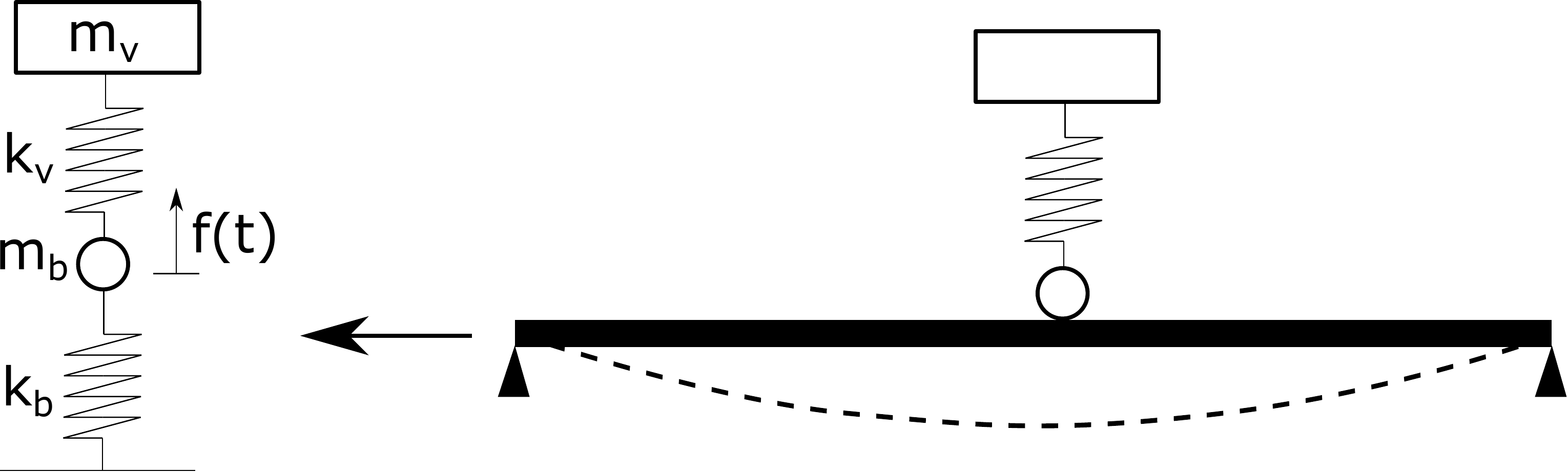}
\caption{schematic of the coupled setup}
\label{fig:proof}
\end{figure}

In this section, a closed-form theoretical proof for validity of the simplified model is presented. Generally, vehicle-bridge interaction is a complex model to be solved in closed-form, however, simplified models can be used for proof of concept \citep{fryba2013vibration,yang2004vehicle}. The objective here is to show that a coupled VBI system subject to external stochastic excitations produces bridge and vehicle responses that are very close to the responses of an uncoupled system, especially if the bridge is long and heavy. For this purpose, the mass and spring system shown in Figure \ref{fig:proof} is considered in which the vehicle is located at the mid-span of the beam with no motion and in full interaction (no damping is considered with no loss of generality). The random spatio-temporal load of the bridge is also lumped into an effective point load that is applied to the bridge mass. In particular, the proof intends to show that the coupling of the bridge response $x_b$ to the vehicle interaction decays as the bridge dimensions grow. 

From Figure \ref{fig:proof}, the beam is modeled as a unidirectional spring, while the vehicle is a single DOF system. The bridge spring represents the first modal stiffness of the beam. The bridge mass is lumped at the contact point of the two components. The setup constitutes a 2 DOF coupled system with the equation of motion shown in Equation \ref{eq:1}. Using this simplified setup, both responses are calculated in closed-form:

\begin{align}\label{eq:1}
    \begin{bmatrix}
    m_b &   0\\
    0   &   m_v\\
    \end{bmatrix}
    \begin{bmatrix}
    \ddot{x}_b\\
    \ddot{x}_v\\
    \end{bmatrix}
    +
    \begin{bmatrix}
    k_b+k_v &   -k_v\\
    -k_v   &   k_v\\
    \end{bmatrix}
    \begin{bmatrix}
    x_b\\
    x_v\\
    \end{bmatrix}
    =
    \begin{bmatrix}
    f(t)\\
    0\\
    \end{bmatrix}
\end{align}

where $m_b$ and $m_v$ are the bridge and vehicle masses, respectively; Also, $k_b$ and $k_v$ are the stiffnesses for two the components. For further calculations, it is assumed that $m_b = \alpha m_v = \alpha m$ and $k_b = \beta k_v = \beta k$ in which $\alpha$ and $\beta$ are bridge to vehicle mass and stiffness ratios, respectively, and  $\alpha > \beta$. Therefore, using relative mass and stiffness ratios, Equation \ref{eq:1} can be states as:

\begin{align}\label{eq:2}
    \begin{bmatrix}
    \alpha m &   0\\
    0   &   m\\
    \end{bmatrix}
    \ddot{X}
    +
    \begin{bmatrix}
    (1+\beta)k &   -k\\
    -k   &   k\\
    \end{bmatrix}
    X
    =
    \begin{bmatrix}
    f(t)\\
    0\\
    \end{bmatrix}
\end{align}

in which $X=[x_b;x_v]$ contains the bridge and vehicle responses, respectively. In order to solve this equation for X, the first step is to decouple it by using modal transformation using eigenvalue analysis shown in Equation \ref{eq:3}.

\begin{align}\label{eq:3}
    det
    \begin{pmatrix}
    (\beta+1)k-\alpha m \omega^2 & -k\\
    -k & k-m \omega^2\\
    \end{pmatrix}
    = ((\beta+1)k-\alpha m \omega^2)(k-m \omega^2)-k^2 = 0
\end{align}

By assuming $\frac{m\omega^2}{k}=\lambda$ and dividing both sides by $k^2$ we have:

\begin{align}\label{eq:4}
    (\beta+1) - (\beta+1)\lambda - \alpha \lambda + \alpha \lambda^2-1=0 \nonumber \\
    \lambda = \frac{\alpha+\beta\pm \sqrt{(\alpha+\beta+1)^2-4\alpha \beta}}{2\alpha}
\end{align}

One can simply assume that $\alpha+\beta+1 \approx \alpha + \beta$ since ratios are significantly large (especially the mass ratio $\alpha$) when considering commercial vehicles and mid- to long-span bridges. This helps further simplifications as shown in Equation \ref{eq:5}:

\begin{align}\label{eq:5}
    &\lambda = \frac{\alpha+\beta\pm \sqrt{(\alpha+\beta)^2-4\alpha \beta}}{2\alpha} = \frac{\alpha+\beta \pm (\alpha-\beta)}{2\alpha} \nonumber \\
    &\lambda_1 = 1 \Rightarrow \omega_1 = \sqrt{\frac{k}{m}} = \omega_v \nonumber \\
    &\lambda_2 = \frac{\beta}{\alpha} \Rightarrow \omega_2 = \sqrt{\frac{\beta}{\alpha}}\omega_v
\end{align}

It is worth noting that from Equation \ref{eq:5}, one of the natural frequencies is equal to the vehicle's fundamental frequency. Once the eigenvalues are found, eigenvectors can be derived to allow for modal superposition. For brevity, this calculation is summarized and the final mode shapes are presented in Equation \ref{eq:6}.

\begin{align}\label{eq:6}
    \Phi=
    \begin{bmatrix}
    \frac{1}{\beta-\alpha+1}    &   \frac{\alpha-\beta}{\alpha}\\
    1   &   1\\
    \end{bmatrix}
    =
    \begin{bmatrix}
   \phi_{11}&  \phi_{12}\\
    \phi_{21}&  \phi_{22}
    \end{bmatrix} 
\end{align}

In Equation \ref{eq:1}, $f(t)$ is the applied load function, which is ultimately assumed as an ambient white noise for a random traffic network (i.e., Gaussian white noise $\sim\mathcal{N}(0,\sigma^2)$). In order to calculate the response of the system to such loads, one approach is to convert it to a sum of sinusoidal waves using Fourier transform. For a Gaussian white noise, the spectral density function is a continuous function of a constant value (the value equals $\sigma^2$). Therefore, for simplicity, the response of the system subject to a single sinusoidal load is found in closed-form and then, the effect of different frequencies is evaluated by parametric study to determine whether the same conclusion is valid over the entire frequency band. Therefore, $f(t) = A_e sin(\omega_e t)$ is defined, in which $A_e$ and $\omega_e$ are the sinusoidal amplitude and frequency, respectively. To convert the equation of motion shown in Equation \ref{eq:1} to modal coordinates, we premultiply both sides by $\Phi^T$. The modal force vector and modal stiffness are then calculated as shown in Equation \ref{eq:7}:

\begin{align}\label{eq:7}
    &\Phi^T F(t) =
    \begin{bmatrix}
    \frac{1}{\beta-\alpha+1}    &   1-\frac{\beta}{\alpha}\\
    1   &   1\\
    \end{bmatrix}^T
    \begin{bmatrix}
    A_e sin(w_e t)\\
    0\\
    \end{bmatrix}
    =
    \begin{bmatrix}
    \frac{A_e}{\beta-\alpha+1}  sin(w_e t)\\
    \frac{A_e(\alpha-\beta)}{\alpha} sin(w_e t)\\
    \end{bmatrix} \nonumber \\
    &\hat{K} = \Phi^T K \Phi =
    \begin{bmatrix}
    \hat{k_1}&  0\\
    0&  \hat{k_2}
    \end{bmatrix}
    =
    \begin{bmatrix}
    \left[\frac{\alpha}{(\beta-\alpha+1)^2}+\frac{\alpha-\beta-2}{\beta-\alpha+1}\right]k&  0\\
    0&  \left[\frac{\beta^3+(1-2\alpha)\beta^2+\beta\alpha^2}{\alpha^2}\right]k
    \end{bmatrix}\nonumber    
    \\
    &x_b = \phi_{11} q_1 + \phi_{21} q_2 \nonumber \\
    &\hat{m_1}\ddot{q_1} + \hat{k_1}q_1 = \frac{A_e}{\beta-\alpha+1}  sin(w_e t) \nonumber \\
    &\hat{m_2}\ddot{q_2} + \hat{k_2}q_2 = \frac{A_e(\alpha-\beta)}{\alpha} sin(w_e t)
\end{align}

The steady-state responses of the single-degree of freedom systems subject to a harmonic load have the following form shown in Equation \ref{eq:8}: 

\begin{align}\label{eq:8}
    &q_1(t) = \frac{\frac{A_e}{\beta-\alpha+1}}{\hat{k_1}}\frac{1}{1-\gamma^2}.sin(\omega_e t) \nonumber \\
    &q_2(t) = \frac{\frac{A_e(\alpha-\beta)}{\alpha}}{\hat{k_2}}\frac{1}{1-\frac{\alpha}{\beta}\gamma^2}.sin(\omega_e t)
\end{align}

in which $\gamma=\omega_e/\omega_v$. For a unit amplitude of the external load (i.e., $A_e=1$) and by substitution of stiffness from Equation \ref{eq:7} to Equations \ref{eq:8}, the harmonic amplitudes are calculated as follows:

\begin{align}\label{eq:9}
    &amp(q_1) = \frac{\beta-\alpha+1}{(\gamma^2-1)(\alpha^2-2\alpha\beta-4\alpha+\beta^2+3\beta+2)k} \nonumber \\
    &amp(q_2) = \frac{\alpha(\alpha-\beta)}{(\beta-\alpha\gamma^2)(\alpha^2-2\alpha\beta+\beta^2+\beta)k}
\end{align}

Finally, by modal superposition of two modal responses, the amplitude of the total harmonic vibration of the bridge is calculated as shown in Equation \ref{eq:10}:

\begin{align}\label{eq:10}
    &amp(x_b) = \phi_{11} \times amp(q_1) + \phi_{21} \times amp(q_2) =  \nonumber \\
    &\frac{1}{k}\left[\frac{1}{(\gamma^2-1)(\alpha^2-2\alpha\beta-4\alpha+\beta^2+3\beta+2)}+\frac{(\alpha-\beta)^2}{(\beta-\alpha\gamma^2)(\alpha^2-2\alpha\beta+\beta^2+\beta)}\right]
\end{align}

\begin{figure}[!ht]
\centering\includegraphics[width=0.5\linewidth]{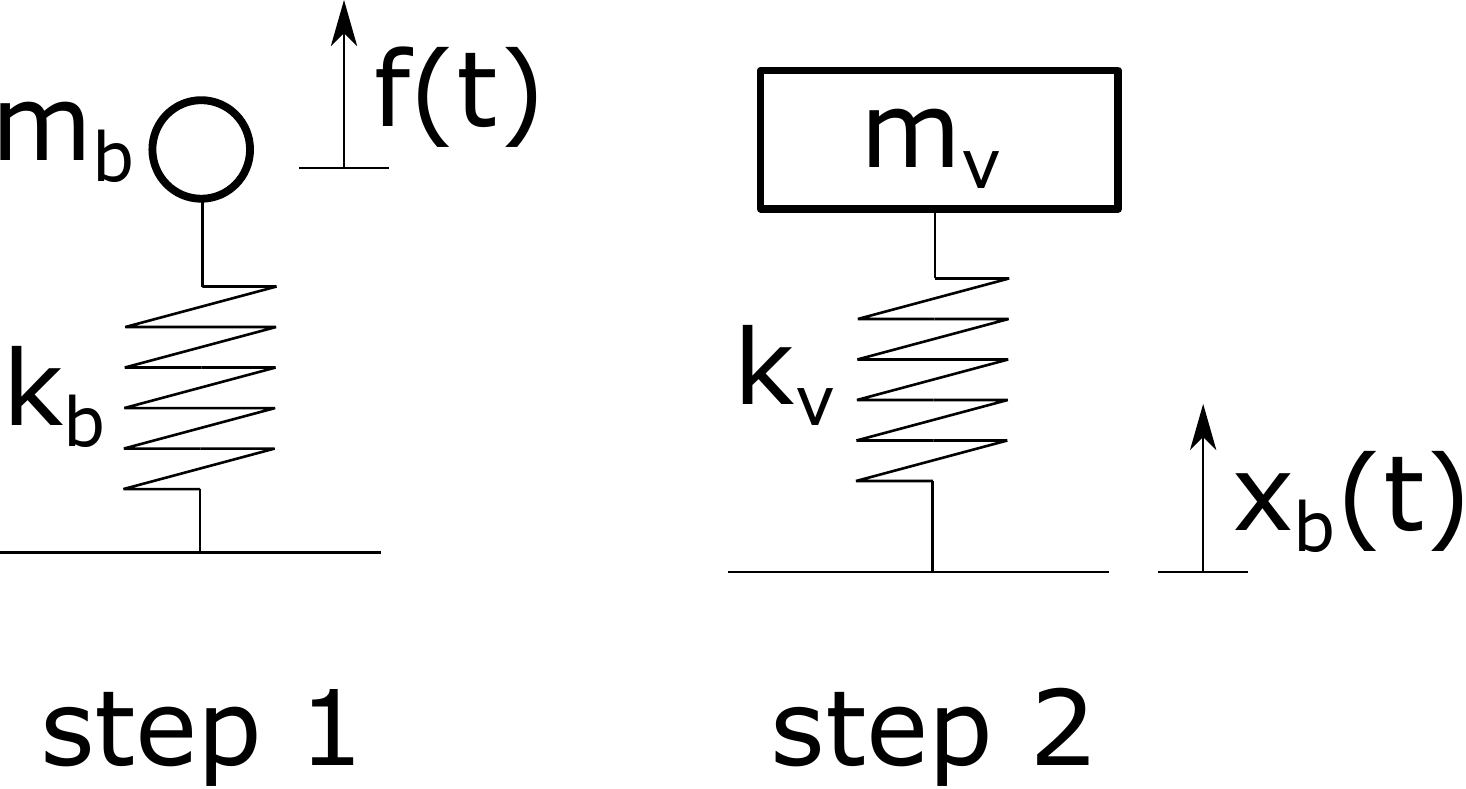}
\caption{schematic of the uncoupled setup}
\label{fig:proof2}
\end{figure}

So far, the bridge response from the fully coupled setup is derived. In order to find the bridge response using the second approach (i.e., the simplified model), the setup shown in Figure \ref{fig:proof2} is assumed. The bridge model is individually subject to the external load and responds to it. The response is then applied to an isolated vehicle model to produce the vehicle response. The closed-form solution for the bridge response in such an uncoupled setup is trivial and shown in Equation \ref{eq:11}.

\begin{align}\label{eq:11}
    &m_b \ddot{x_b} + k_b x_b=A_e sin(\omega_e t) \nonumber \\
    &x_b = \frac{A_e}{k_b}.\frac{1}{1-\frac{\omega_e}{\omega_b}^2}.sin(\omega_e t) \nonumber \\
    &amp(x_b) = \frac{1}{k(\beta+\alpha\gamma^2)}
\end{align}

\begin{figure}[!ht]
\centering\includegraphics[width=0.7\linewidth]{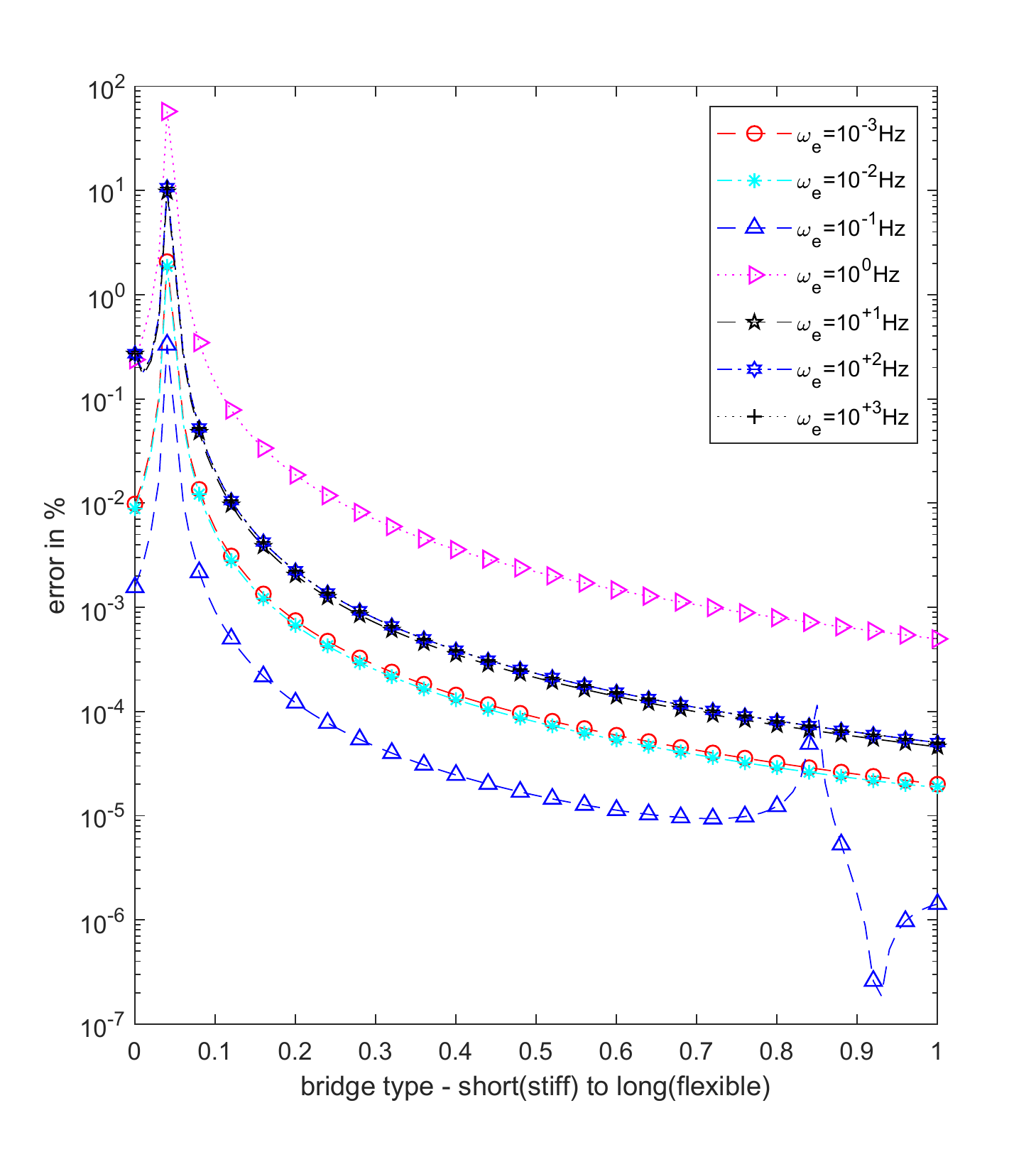}
\caption{Results of the theoretical approach: parametric study shows the extent of the error for different bridge types and loading frequencies when using the simplified bridge-vehicle simulation approach.}
\label{fig:proof_parametric}
\end{figure}

Once Equations \ref{eq:10} and \ref{eq:11} are derived, the parametric study can take place. Both equations are functions of $\alpha$, $\beta$, and $\gamma$. By plotting the response error between these two solutions for different ranges of these three parameters, the extent of the error in the simplified decoupled model can be investigated. Intuitively, as the bridge size increases, the stiffness of the structure decreases (i.e., longer bridges are more flexible), and the mass increases, resulting lower fundamental frequencies. The main objective is to observe the sensitivity of the error to the bridge size. Therefore, different mass and stiffness ratio pairs are plugged into both equations and errors are calculated. In addition, different loading frequencies are also examined. The mass and stiffness ratios ($\alpha$ and $\beta$) used for this purpose range $[50:10,000]$ and $[500:10]$, respectively, modeling short (stiff) bridges to long (flexible) ones. Loading frequencies spread exponentially from $10^{-3}$Hz to $10^3$Hz to envelope a sufficiently wide range of loading frequencies. Figure \ref{fig:proof_parametric} summarizes the outcomes of the parametric study. Note that the x axis corresponds to different mass and stiffness ratio pairs, which is normalized to better convey the qualitative aspect of the plot (i.e., $0$ is the stiffest bridge while $1$ stands for the most flexible one). 

Figure \ref{fig:proof_parametric} demonstrates that based on the closed-form solutions, what would be the extent of error in the simplified simulation method for different types of bridges. As the bridge size increases, the error between two methods decays substantially (e.g., below $0.1\%$ error for long bridges). This supports the idea that an uncoupled simplified solution is accurate enough when the bridge length increases. The figure also shows that there is a range of bridges in which the error is not negligible (for relatively short bridges the error can be up to $50\%$ when the loading frequency resonate with the natural frequency of the vehicle). Also notice that the same trend occurs for different loading frequencies, with maximum error near the vehicle resonance frequency. \par

In this part, using our simplified model we showed that the uncoupled simulation approach yields accurate results when compared to the fully coupled approach, especially when the bridge size grows. In the next section, the results from a more detailed numerical simulation of the vehicle-bridge interaction are presented in order to incorporate other aspects of the VBI problems, such as vehicle motions and road roughness profile. 

\section{Numerical Analysis}\label{sec:numeric}

In this section, the VBI problem is modeled numerically in Matlab and the results are compared with the signals from the simplified simulation approach. In this numerical case study, six bridges with different span lengths are modeled in SAP2000 and two simulation approaches are implemented. The exact numerical approach for modeling the bridge response interacting with a moving vehicle (roughness included) is adopted from \cite{gonzalez2012identification} as presented in Algorithm \ref{alg:conventional}.

\begin{figure}[!ht]
\centering\includegraphics[width=0.5\linewidth]{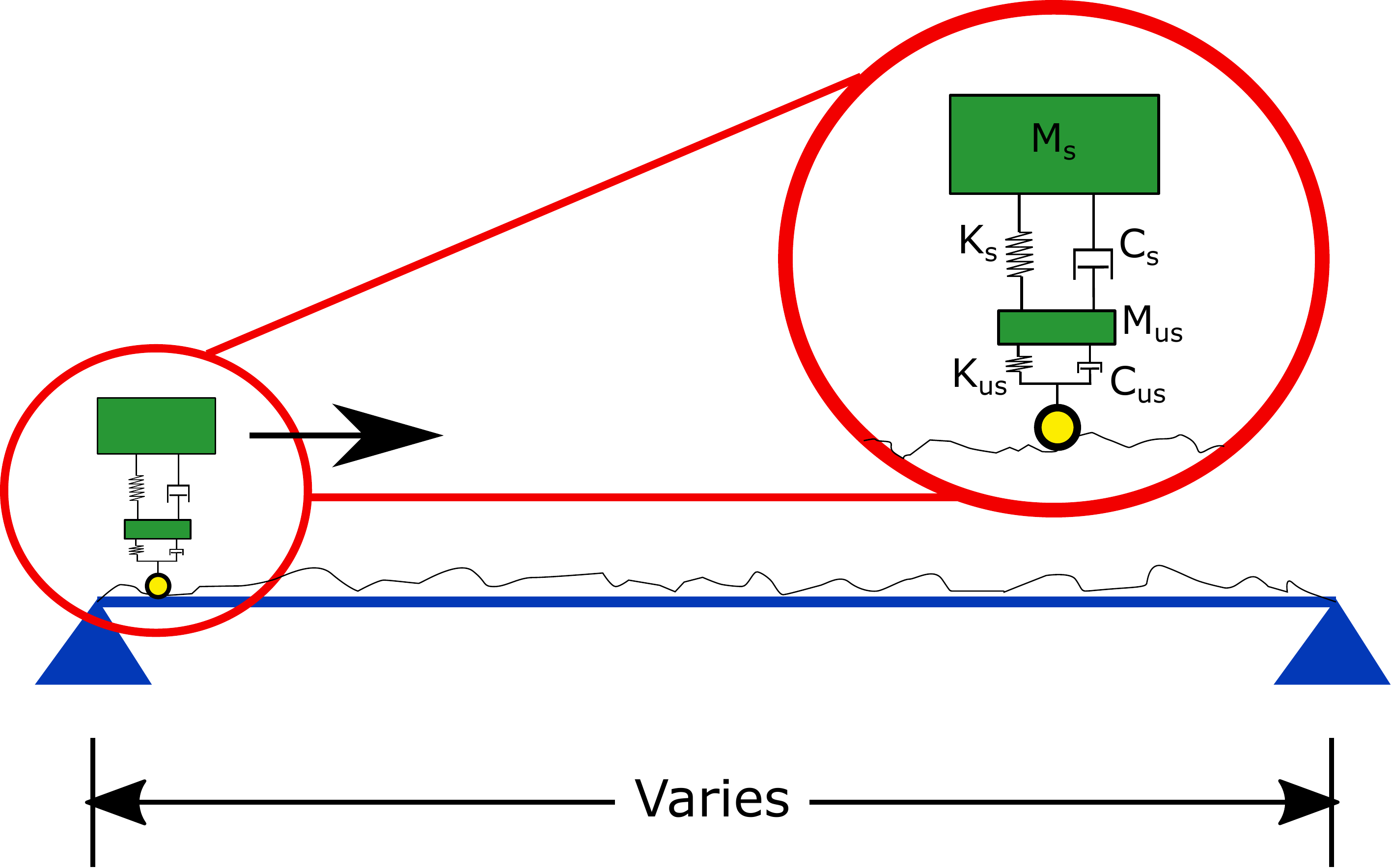}
\caption{Schematic of simulated model - roughness profile is also included}
\label{fig:simul_scheme}
\end{figure}

The bridge setup is shown in Figure \ref{fig:simul_scheme}. The span varies from 15 m (very short and stiff bridge) to 500 m (long and flexible bridge), with mechanical properties shown in Table \ref{tbl:beam_props}. The bridge is 3D modeled in SAP2000 using prismatic beams with box cross-sections. The structural behavior is assumed linearly elastic for consistency with operational modal analysis. The road roughness profile is adapted according to ISO standard for a road class 'A' \citep{organizacion1995mechanical} which is the case for a well maintained highway road condition. At each time instance, the bridge model is analyzed dynamically using Newmark-$\beta$ method using matrices imported from SAP2000. For the vehicle, first a quarter-car model is adopted with the properties shown in Table \ref{tbl:vcl_props}. This vehicle simulates suspension properties of a commercial vehicle with high damping and low natural frequency (which are critical factors for a comfortable ride \citep{milliken2002chassis}). The second vehicle is a quarter-car model of a heavy truck adopted from \citep{harris2007reduction,elhattab2016drive} with properties  shown in Table \ref{tbl:trk_props}. The second vehicle is selected to investigate the approximation error of using the simplified method for heavy sensing agents when the weight is not negligible.\par 

\begin{table}[!h]
\centering
\caption{Bridge spans and cross-section dimensions}
\begin{tabular}{lcccccc}
\toprule
Span length {[}m{]}                                                    & \textit{15m} & \textit{30m} & \textit{50m} & \textit{100m} & \textit{200m} & \textit{500m} \\ \midrule
Outside depth {[}m{]}                                                  & 0.60          & 1.10          & 1.60          & 2.40           & 3.00             & 5.00             \\
Outside width {[}m{]}                                                  & 0.3          & 0.50          & 1.30          & 2.00             & 2.50           & 4.00             \\
Flange thickness {[}m{]}   & 0.04         & 0.05         & 0.10          & 0.15          & 0.15          & 0.50           \\
Web thickness {[}m{]}                                                  & 0.02         & 0.03         & 0.05         & 0.10           & 0.10           & 0.25          \\
Fundamental freq. {[}Hz{]} 
& 8.03         & 3.63         & 2.05         & 0.75          & 0.24          & 0.06          \\ \bottomrule
\end{tabular}
\label{tbl:beam_props}
\end{table}

\begin{table}[!h]
\parbox{.45\linewidth}{
\centering
\caption{Commercial vehicle properties}
\begin{tabular}{l l l}
\toprule
\textit{Property Name} & \textit{Value} & \textit{Units}\\
\midrule
Unsprung Mass & 69.9 & Kg \\
Sprung Mass & 466.0 & Kg \\
Tire Damping & 0.0 & Ns/m \\
Suspension Damping & 2796.0 & Ns/m \\
Tire Stiffness & 3043.0 & N/m \\
Suspension Stiffness & 290.3 & N/m \\
Fundamental Frequency & 1.2 & Hz \\
\bottomrule
\end{tabular}
\label{tbl:vcl_props}
}
\parbox{.45\linewidth}{
\centering
\caption{Heavy truck properties}
\begin{tabular}{l l l}
\toprule
\textit{Property Name} & \textit{Value} & \textit{Units}\\
\midrule
Unsprung Mass & 700.0 & Kg \\
Sprung Mass & 17,300.0 & Kg \\
Tire Damping & 0.0 & Ns/m \\
Suspension Damping & $1.0\times 10^4$ & Ns/m \\
Tire Stiffness & $1.75\times 10^6$ & N/m \\
Suspension Stiffness & $4.0\times 10^5$ & N/m \\
Fundamental Frequency & 0.69 & Hz \\
\bottomrule
\end{tabular}
\label{tbl:trk_props}
}
\end{table}

For a fair comparison, the vehicle's speed is kept constant among all bridge spans ($10m/sec$). Finally, the traffic load is modeled as a random ambient load uniformly applied over the span with the amplitude proportional to the number of vehicles. In particular, for $n$ vehicles, a random and sparse matrix is generated in which the sum of forces in each row (i.e., for each time instance) is equal to $n\times2,000\times g$ N, assuming $2,000$ kg for the average weight of a commercial vehicle and $g$ is the gravity acceleration. Four traffic levels are considered for each span length with $n=0,10,20,50$ ($n=0$ models an isolated bridge while $n=50$ models a bridge with 50 vehicles moving while being scanned by the sensing agent). The bridge is modeled as a MDF system with $0.1m$ spatial discretization (e.g., 15 m long bridge is modeled with 150 DOFs). For simulating responses using the decoupled model, Algorithm \ref{alg:simplified} is adopted: the random traffic load is firstly applied to the bridge with no consideration for the sensing vehicle. The bridge responses at the vehicle locations are then aligned in space and applied to the model of the sensing vehicle. The vehicle processes the input through its dynamical model (shown in Tables \ref{tbl:vcl_props} and \ref{tbl:trk_props}) and produce the vehicle response. 

\begin{figure}[!h]
\centering
    \begin{subfigure}[t]{0.45\textwidth}
        \centering
        \includegraphics[height=2.1in]{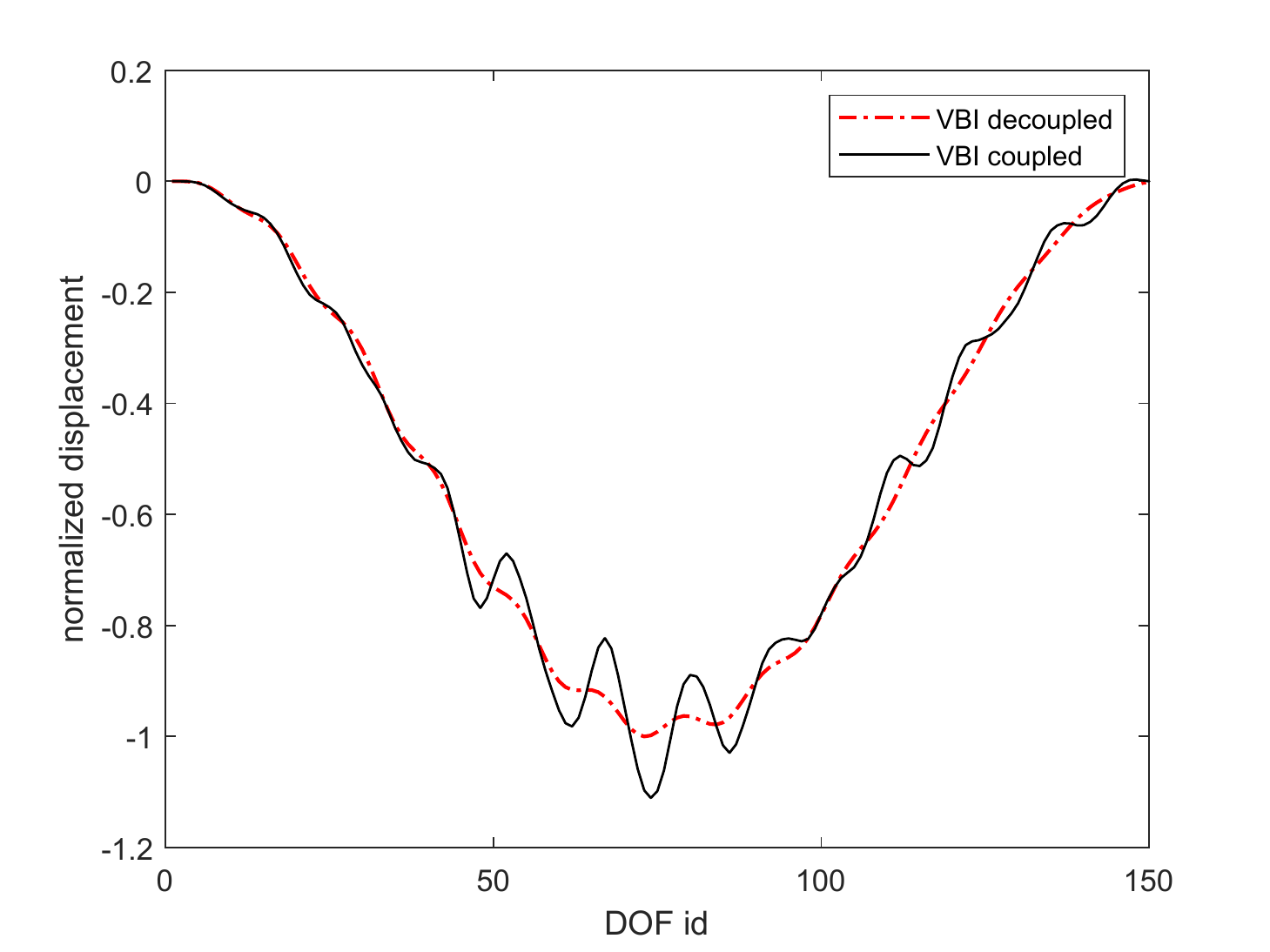}
        \caption{15 m bridge}
    \end{subfigure}%
    ~
    \begin{subfigure}[t]{0.45\textwidth}
        \centering
        \includegraphics[height=2.1in]{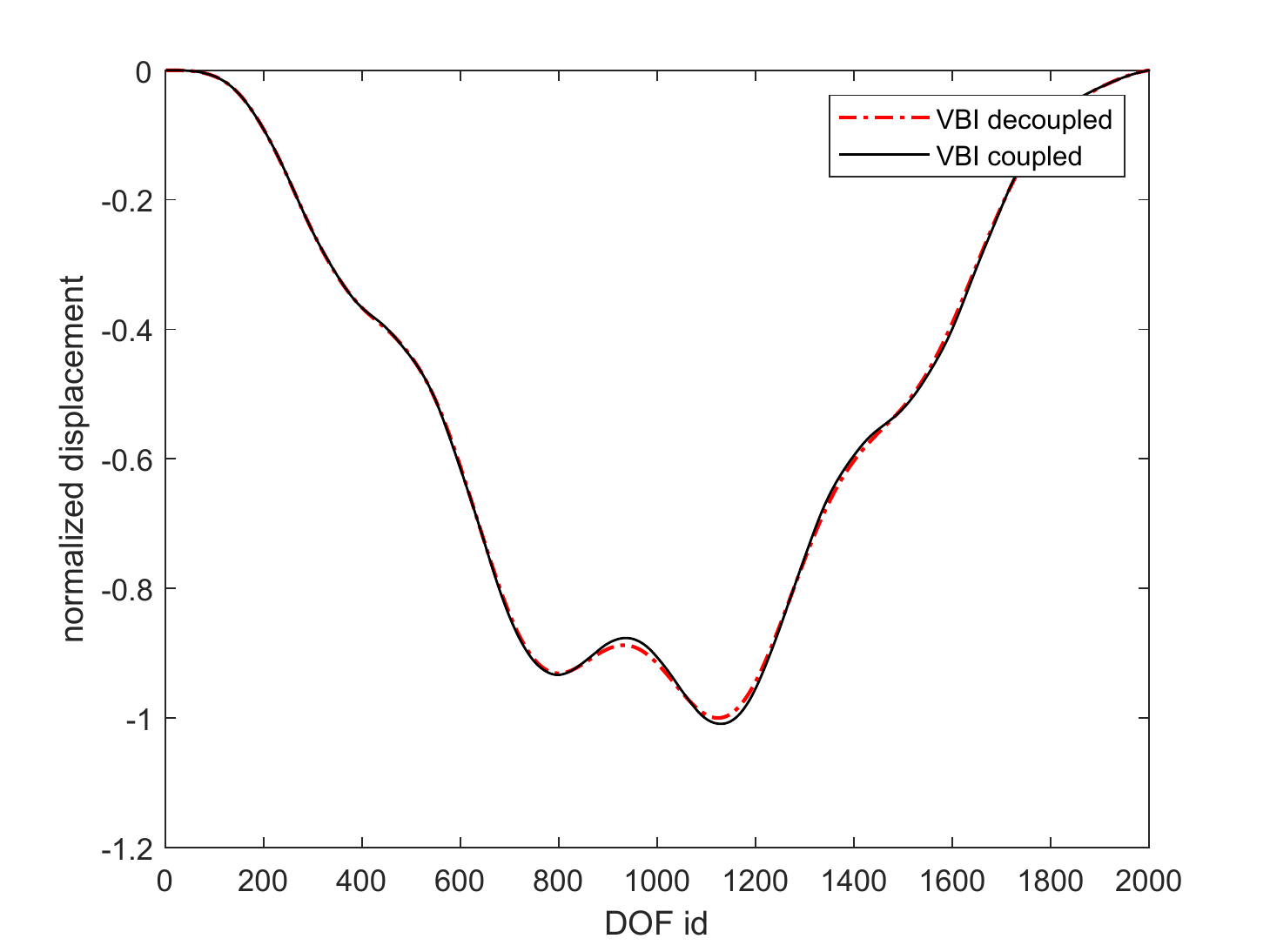}
        \caption{200 m bridge}
    \end{subfigure}%
\caption{Bridge displacement simulation results for the commercial vehicle}
\label{fig:Bridge_disp_vcl}
\end{figure}

\begin{figure}[!h]
\centering
    \begin{subfigure}[t]{0.45\textwidth}
        \centering
        \includegraphics[height=2.1in]{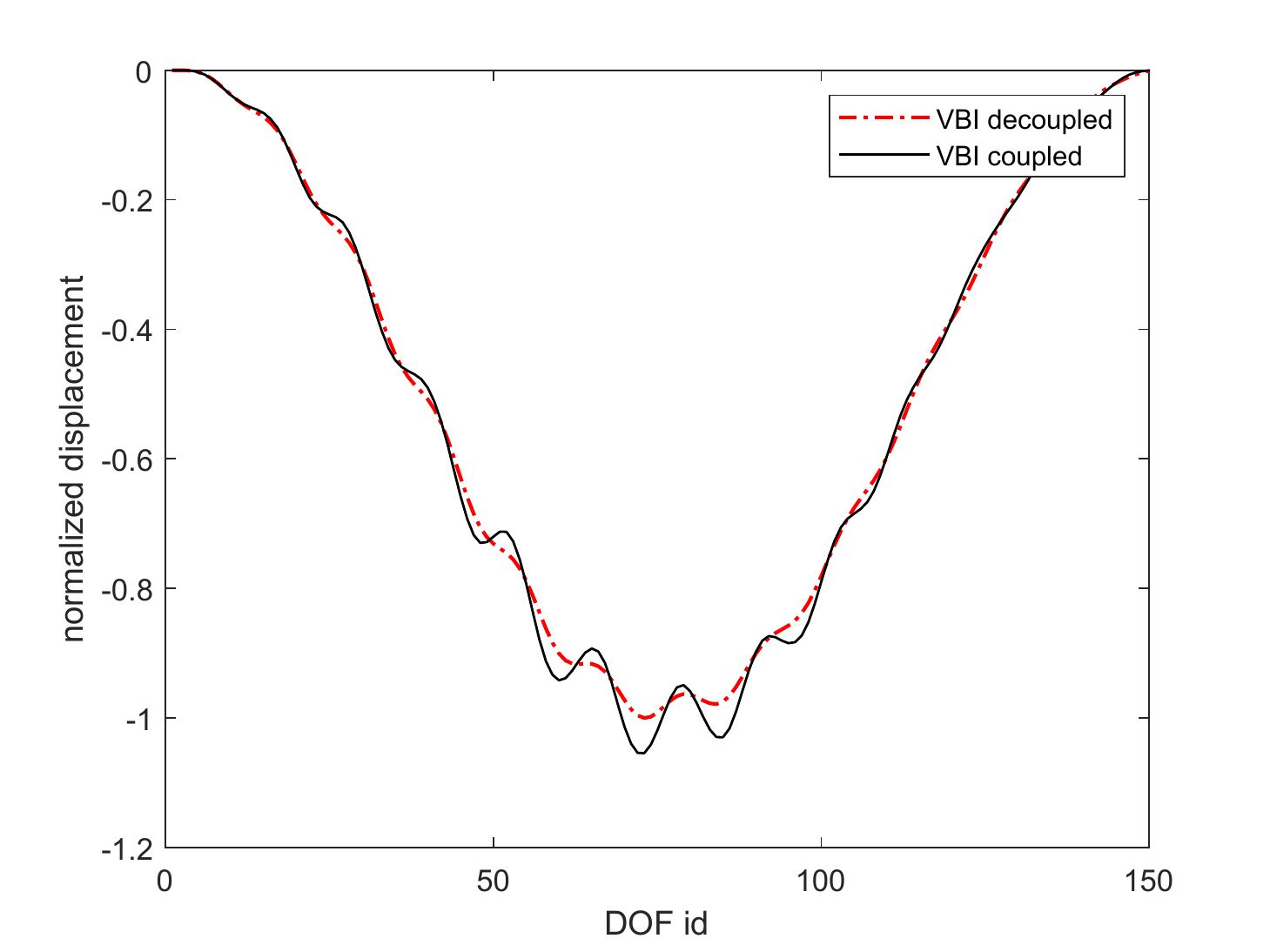}
        \caption{15 m bridge}
    \end{subfigure}%
    ~
    \begin{subfigure}[t]{0.45\textwidth}
        \centering
        \includegraphics[height=2.1in]{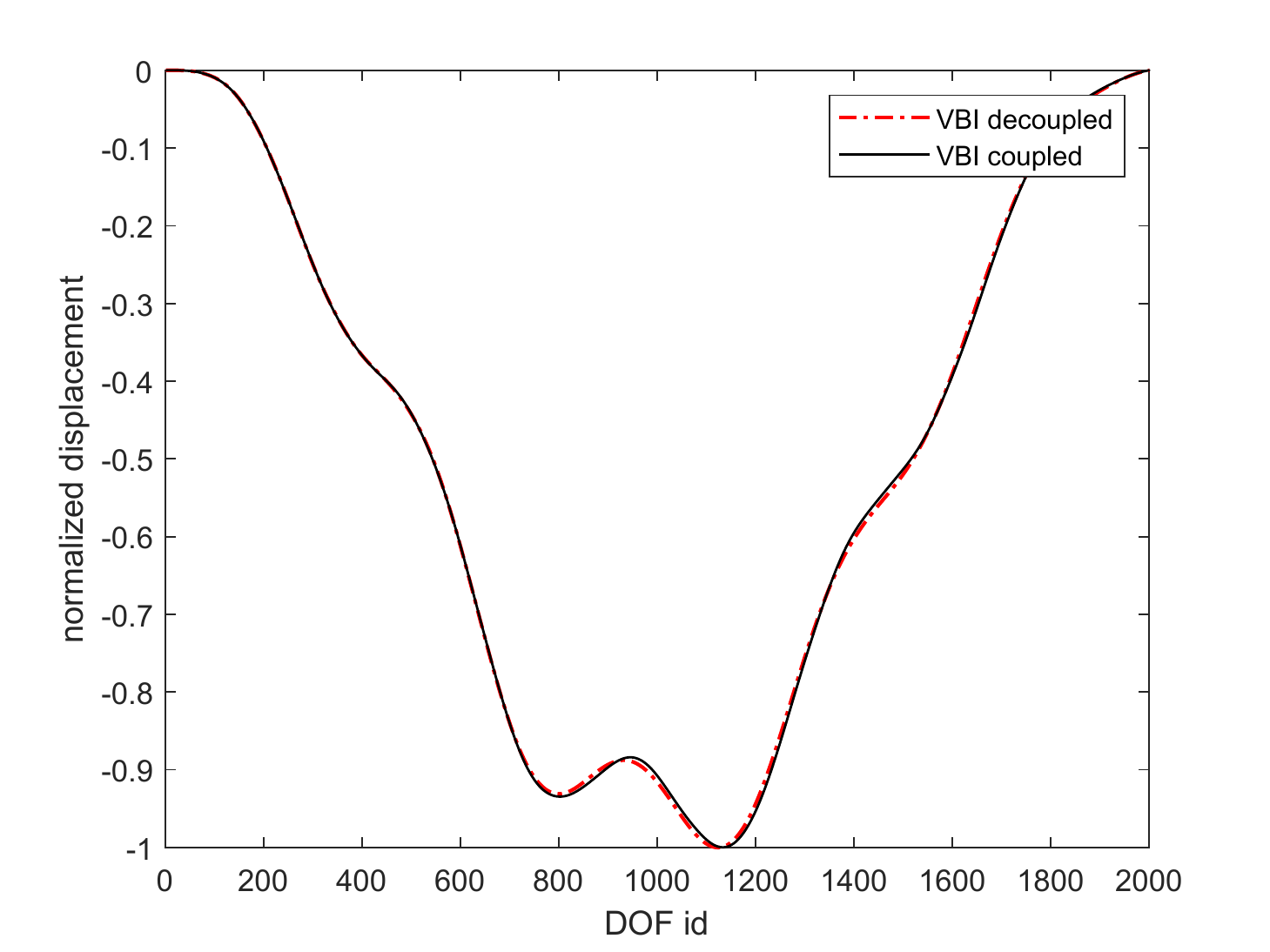}
        \caption{200 m bridge}
    \end{subfigure}%
\caption{Bridge displacement simulation results for the heavy truck}
\label{fig:Bridge_disp_trk}
\end{figure}

The performance of the simplified model is evaluated in terms of the bridge response as well as the vehicle response. From Section \ref{sec:theory} it is expected that the simplified model yield more accurate response estimations as the length of the bridge span increases. For the conventional simulation approach, the acceptance threshold for the bridge response is set to $1.5\times 10^{-12}$ m. For each bridge span and traffic level pairs, bridge and vehicle response signals are simulated using two approaches (in total 24 runs for each vehicle); and the errors between two signals are measured in time and frequency domains using the mean squared error (MSE). For more consistency, the responses are scaled by the absolute maximum values of the displacement signals found from the conventional method. \par

\begin{figure}[!ht]
\centering
    \begin{subfigure}[t]{0.45\textwidth}
        \centering
        \includegraphics[height=2.1in]{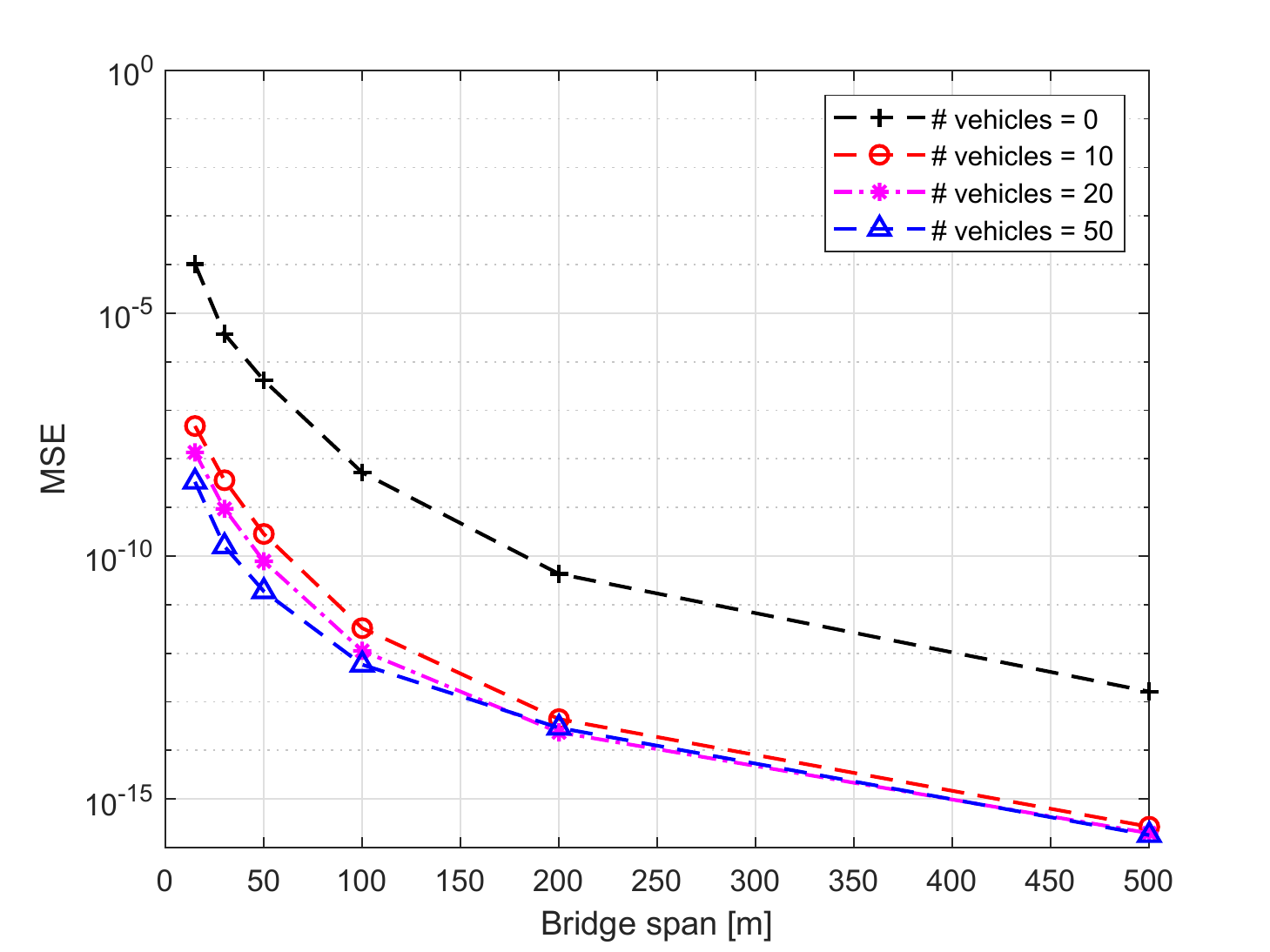}
        \caption{Time signal comparison}
    \end{subfigure}%
    ~
    \begin{subfigure}[t]{0.45\textwidth}
        \centering
        \includegraphics[height=2.1in]{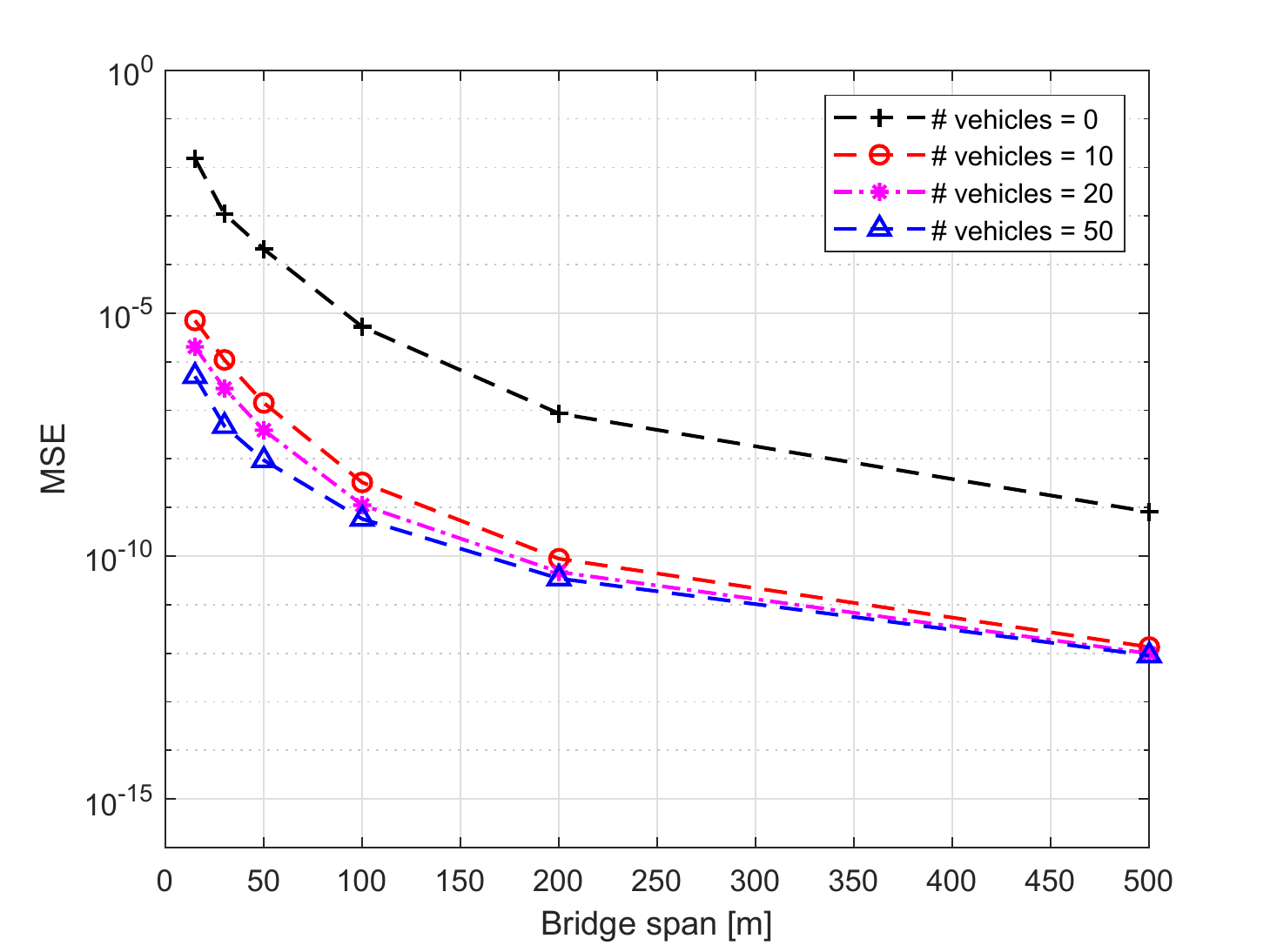}
        \caption{Frequency comparison}
    \end{subfigure}%
\caption{Bridge response comparison for the commercial vehicle in terms of the MSE: The trends show more accurate simulation results as bridge span or traffic volume increases.}
\label{fig:MSE_Bridge_Rgh}
\end{figure}

\begin{figure}[!ht]
\centering
    \begin{subfigure}[t]{0.45\textwidth}
        \centering
        \includegraphics[height=2.1in]{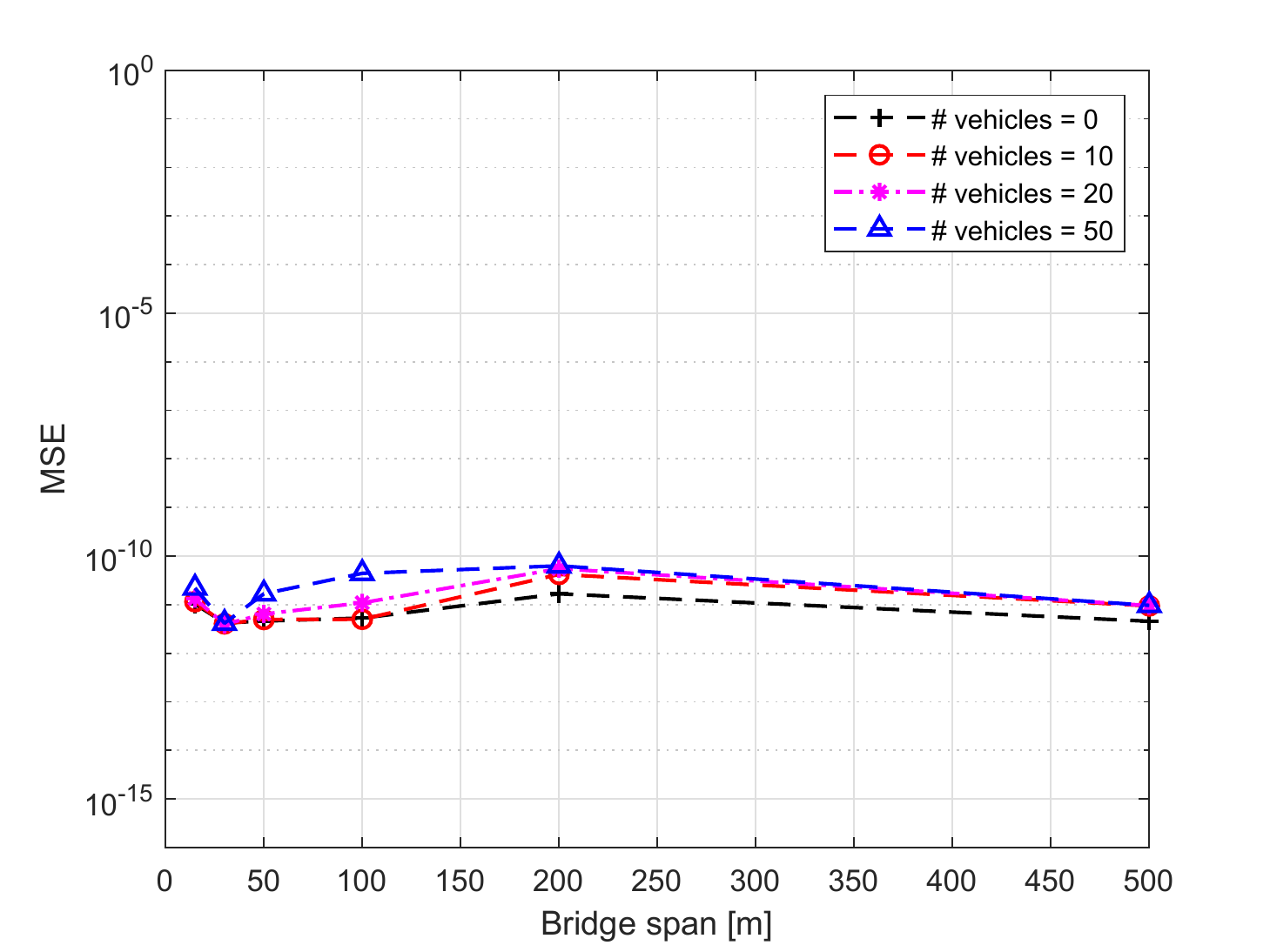}
        \caption{Time signal comparison}
    \end{subfigure}%
    ~
    \begin{subfigure}[t]{0.45\textwidth}
        \centering
        \includegraphics[height=2.1in]{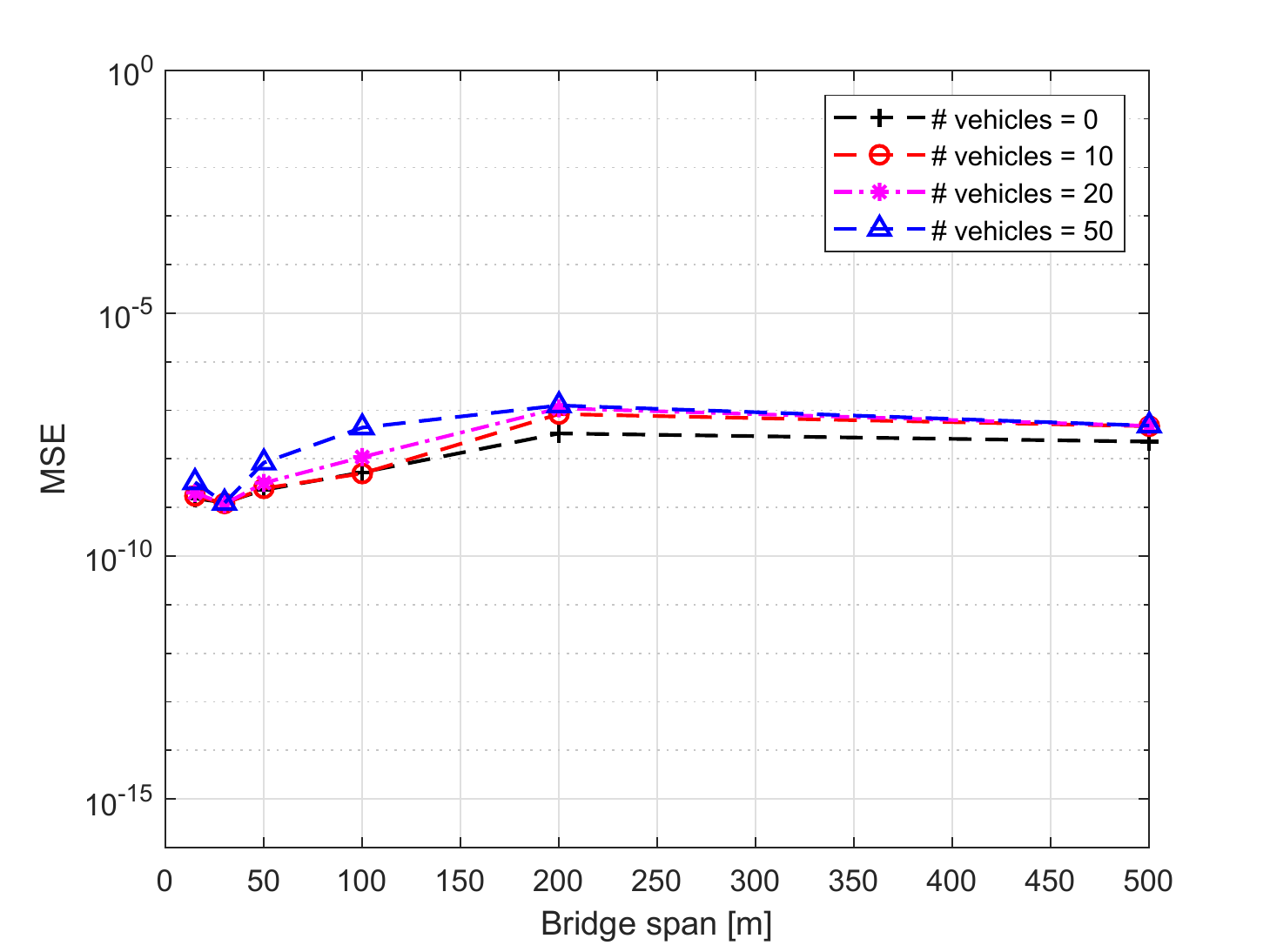}
        \caption{Frequency comparison}
    \end{subfigure}%
\caption{Vehicle response comparison for the commercial vehicle in terms of the MSE: The trends show invariance to the span and the traffic level.}
\label{fig:MSE_Vehicle_Rgh}
\end{figure}

\begin{figure}[!ht]
\centering
    \begin{subfigure}[t]{0.45\textwidth}
        \centering
        \includegraphics[height=2.1in]{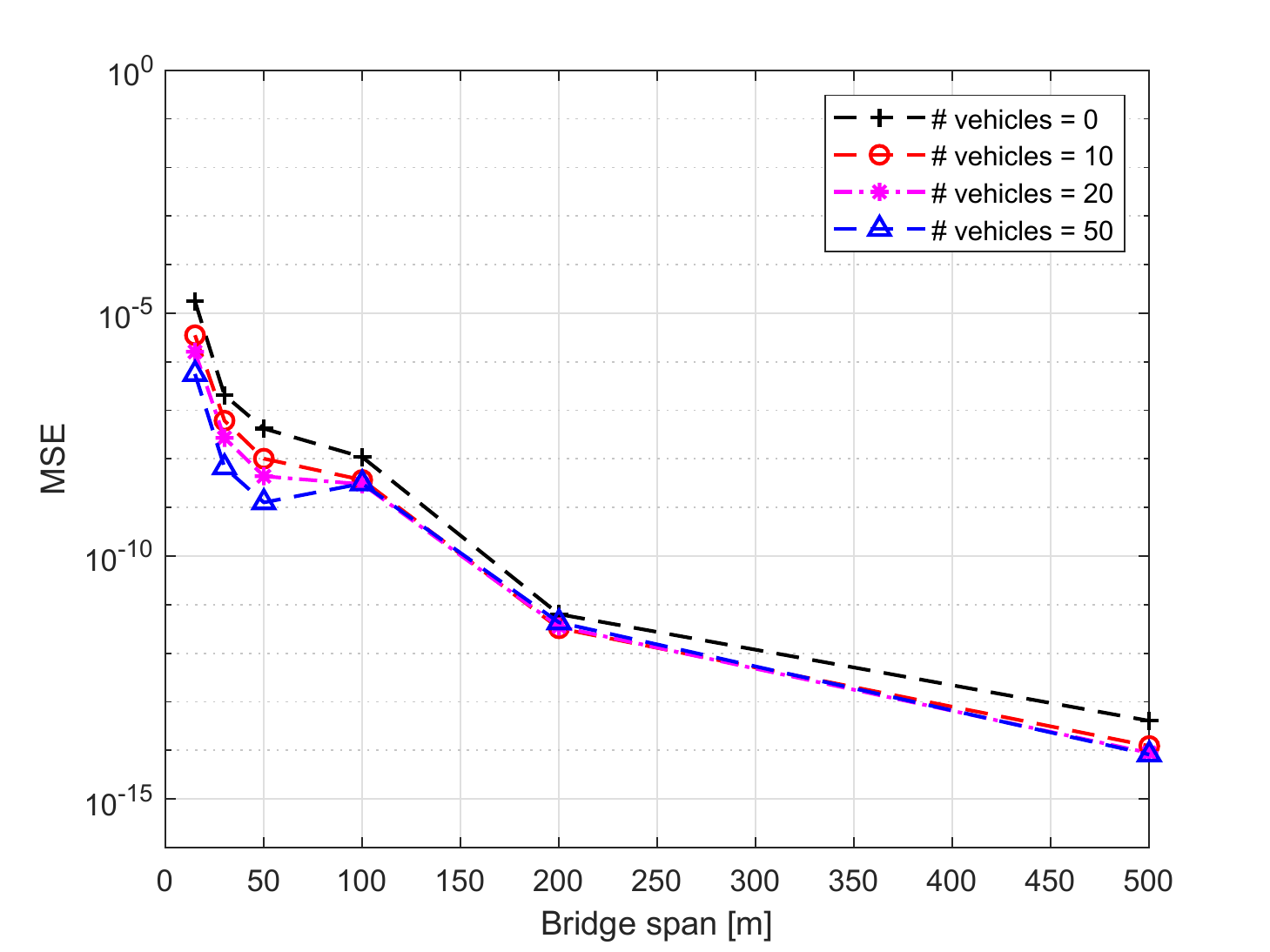}
        \caption{Time signal comparison}
    \end{subfigure}%
    ~
    \begin{subfigure}[t]{0.45\textwidth}
        \centering
        \includegraphics[height=2.1in]{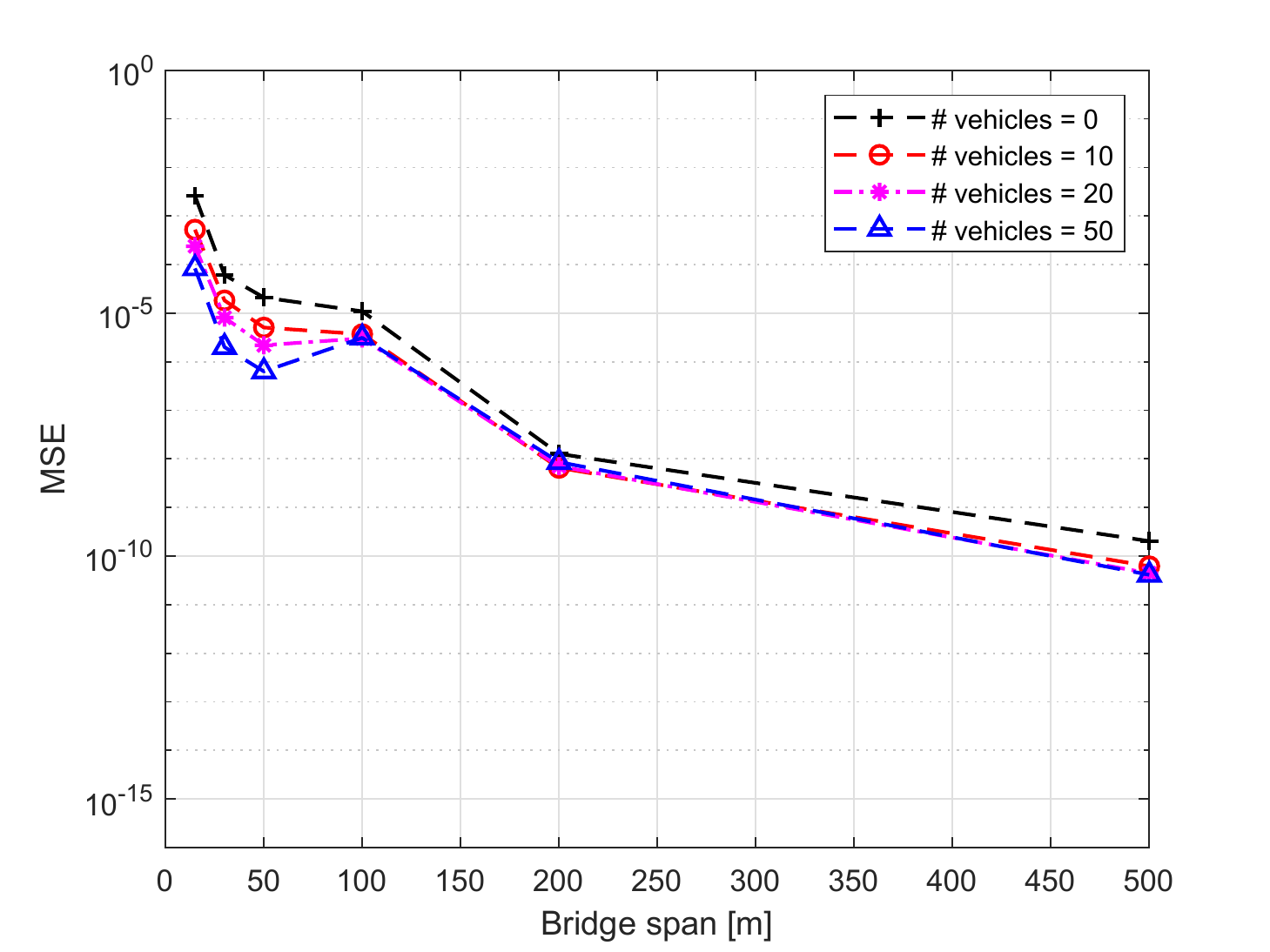}
        \caption{Frequency comparison}
    \end{subfigure}%
\caption{Bridge response comparison for the heavy truck in terms of the MSE: The trends show more accurate simulation results as the bridge span or the traffic volume increases.}
\label{fig:MSE_Bridge_Rgh_heavy}
\end{figure}

\begin{figure}[!ht]
\centering
    \begin{subfigure}[t]{0.45\textwidth}
        \centering
        \includegraphics[height=2.1in]{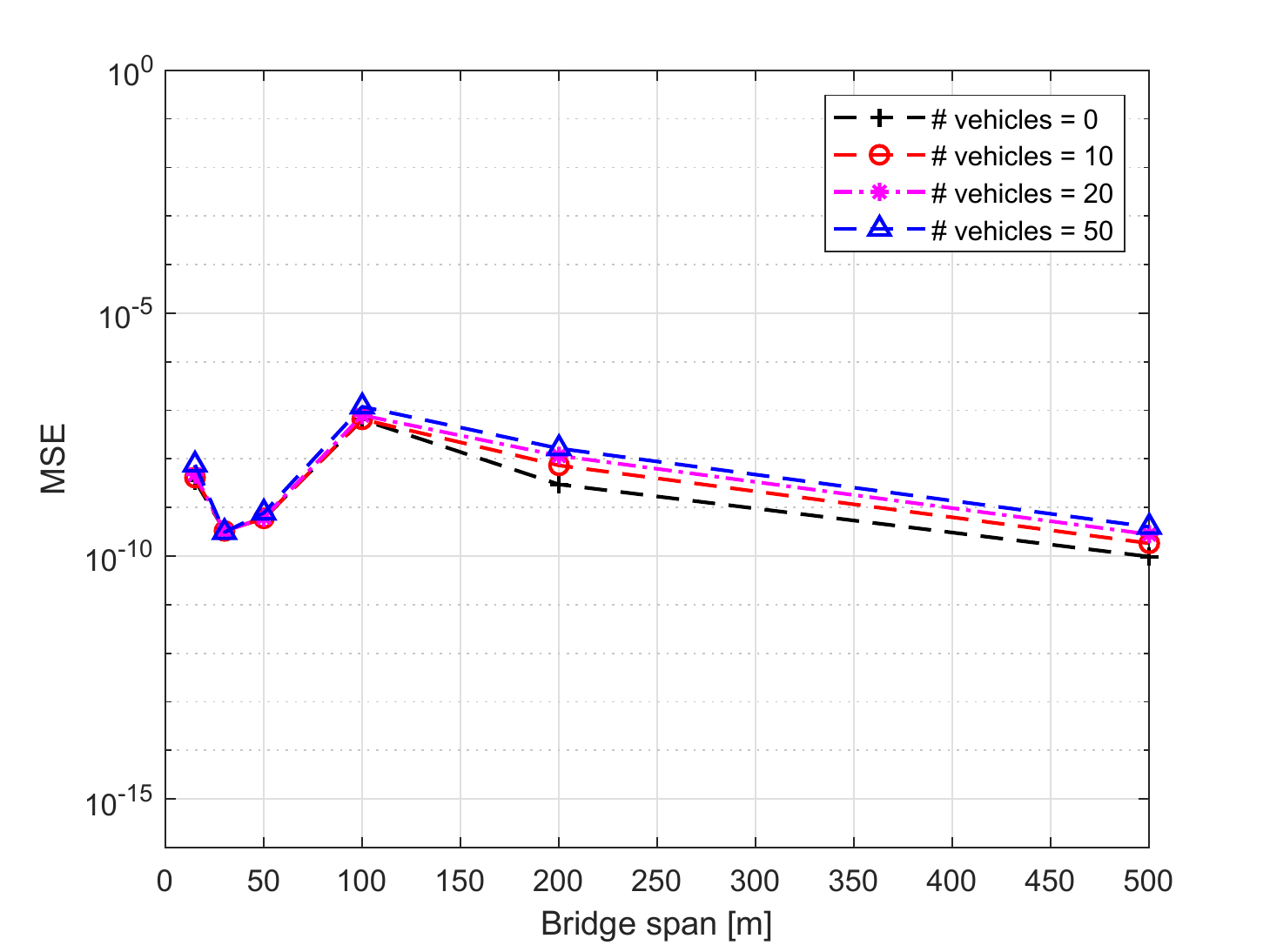}
        \caption{Time signal comparison}
    \end{subfigure}%
    ~
    \begin{subfigure}[t]{0.45\textwidth}
        \centering
        \includegraphics[height=2.1in]{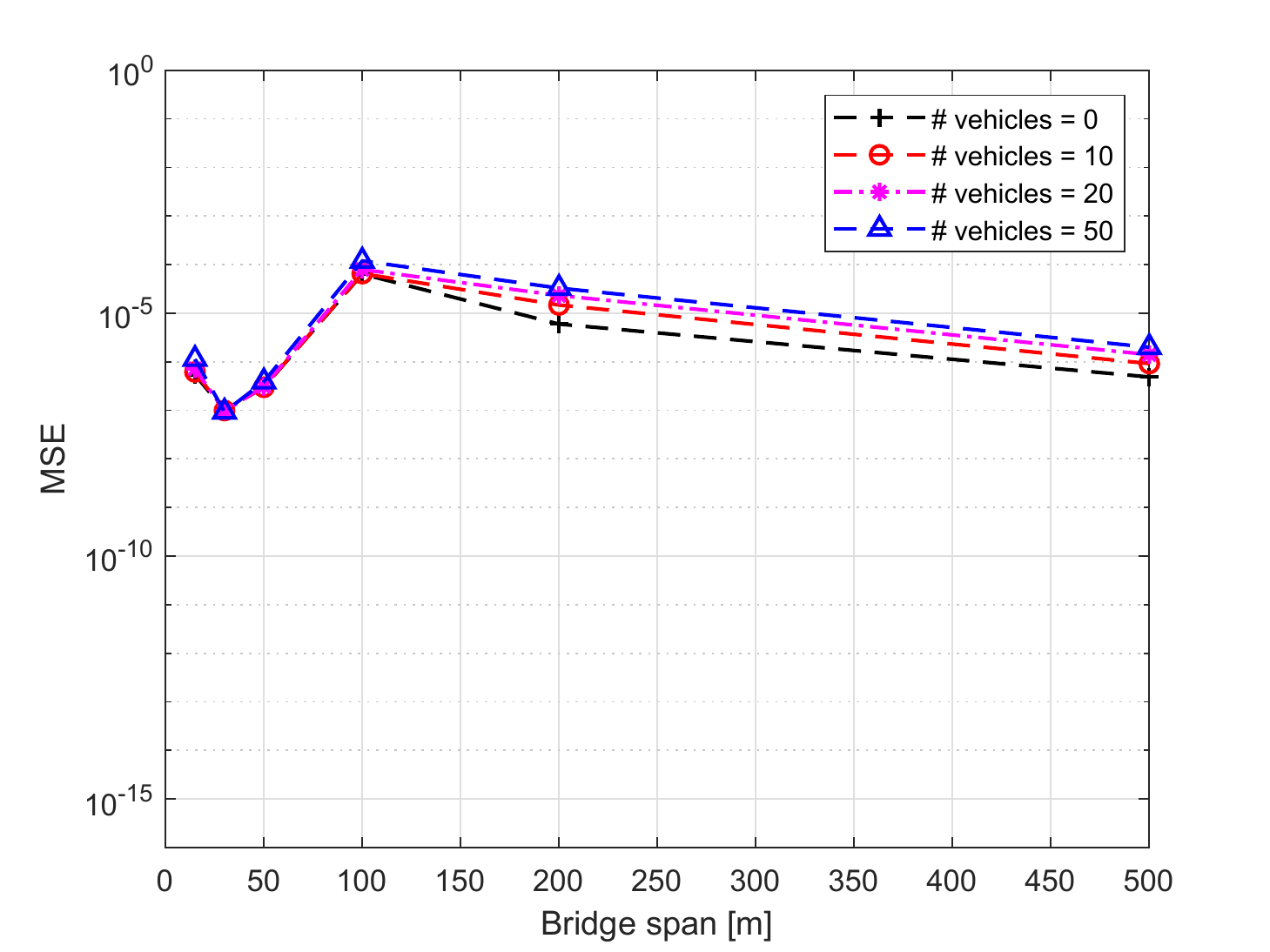}
        \caption{Frequency comparison}
    \end{subfigure}%
\caption{Vehicle response comparison for the heavy truck in terms of the MSE: The trends show that the error peaks when the bridge and the vehicle have close fundamental frequency values.}
\label{fig:MSE_Vehicle_Rgh_heavy}
\end{figure}

Simulated displacement signals for two spans (15 m and 200 m) are shown in Figures \ref{fig:Bridge_disp_vcl} and \ref{fig:Bridge_disp_trk}. For both vehicle types, the bridge response differs noticeably between the conventional and simplified VBI simulations for the 15 m bridge. However, as expected from Section \ref{sec:theory}, as the bridge length increases, the discrepancy between two simulation approaches shrinks in bridge response estimation. The MSE values versus bridge length are also presented in Figures \ref{fig:MSE_Bridge_Rgh} and \ref{fig:MSE_Vehicle_Rgh} for the commercial vehicle and Figures \ref{fig:MSE_Bridge_Rgh_heavy} and \ref{fig:MSE_Vehicle_Rgh_heavy} for the heavy truck to further quantify this observation. Figures \ref{fig:MSE_Bridge_Rgh} and \ref{fig:MSE_Bridge_Rgh_heavy} (error in the bridge response simulations) show a strictly decreasing MSE value as the bridge length increases. In addition, in both cases, as the traffic level increases (i.e., from $n=0$ to $n=50$), the estimation error reduces. This is more evident for the commercial vehicle. Note that the same patterns are deduced from the frequency representation plots. \par

Figures \ref{fig:MSE_Bridge_Rgh} and \ref{fig:MSE_Bridge_Rgh_heavy} show the extent of error for simulating stationary sensors' data that are attached to the bridge. However, what a mobile sensing agent records while scanning the bridge, is not the bridge pure vibrations, but the vehicle response to it. Therefore, Figures \ref{fig:MSE_Vehicle_Rgh} and \ref{fig:MSE_Vehicle_Rgh_heavy} show the accuracy of the vehicle response subject to the bridge motion when comparing the simplified model with the conventional approach. In this case, two sensing agents (i.e., the commercial vehicle versus the heavy truck) react differently. For the commercial vehicle, the responses are relatively insensitive to the span and traffic level and the errors are consistently low for all cases. However, from Figure \ref{fig:MSE_Vehicle_Rgh_heavy}, the truck response is simulated less accurately when the bridge span grows from 15 m to 100 m (for longer bridges, a decaying error trend is observed again). In particular, the frequency estimation error for the heavy vehicle crossing a 100 m long bridge is quite noticeable when using the simplified model. From Table \ref{tbl:trk_props}, the fundamental frequency of the truck is $0.69$Hz which is near resonance for the 100 m long bridge (from Table \ref{tbl:beam_props}, $f=0.75$Hz). Moreover, the vehicle weight is significant, which results in higher interaction forces applied to the bridge and the vehicle itself. In fact, this case highlights that when the bridge and the vehicle have near resonance frequencies, the simplified model works more accurately when the vehicle is lightweight. To validate this, the properties from Table \ref{tbl:trk_props} are downscaled by a factor of 5 (i.e., the same natural frequency while being lighter) and simulation for 100 m long bridge is repeated. The MSE value for $n=50$ from $1.19\times 10^{-4}$ reduced to $5.46\times 10^{-6}$. 

\section{Computational Cost Evaluation}

\begin{figure}[!h]
\centering\includegraphics[width=0.6\linewidth]{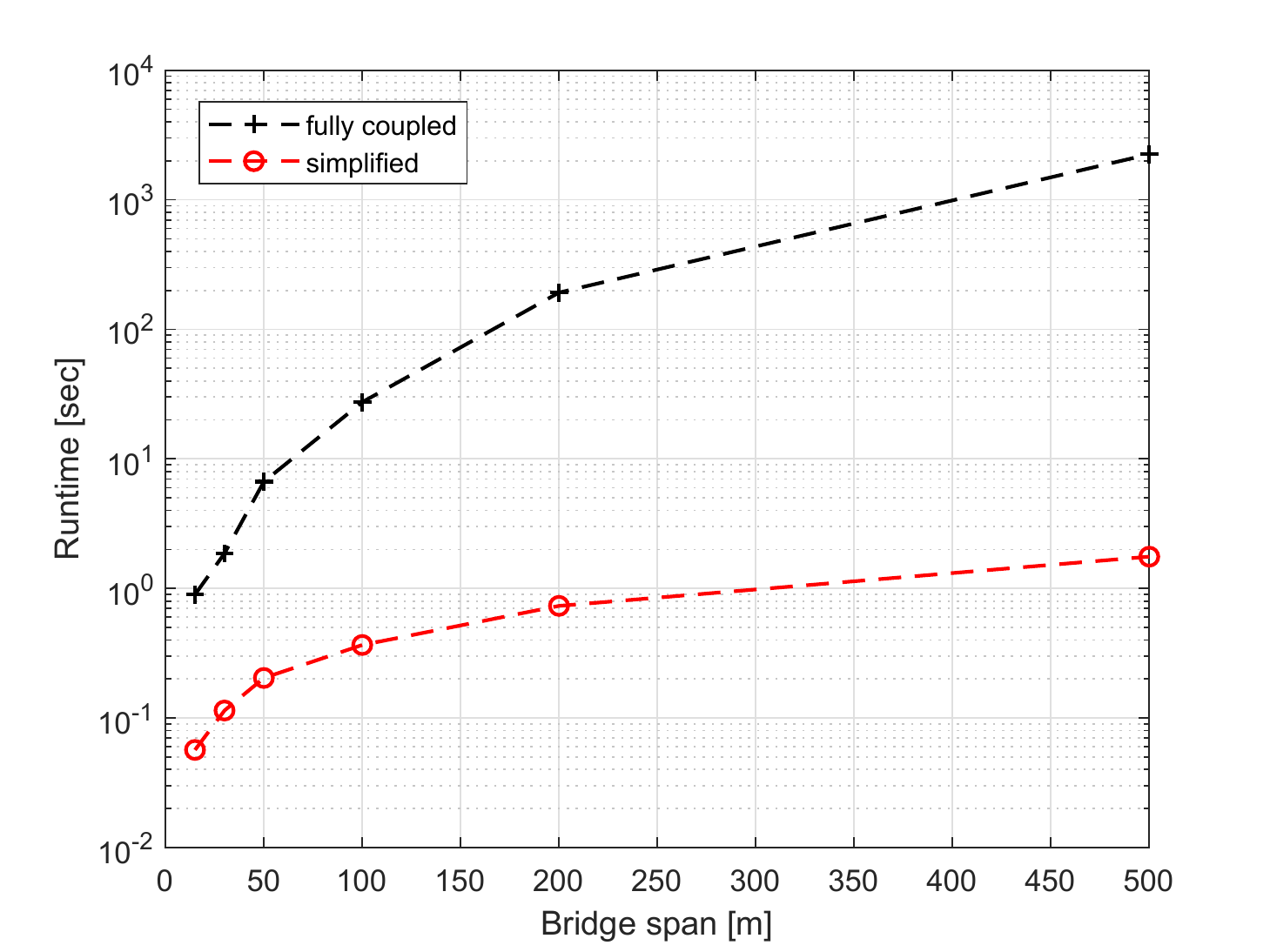}
\caption{Runtime comparison between the conventional and simplified models: the conventional approach is computationally >1,000x slower than the simplified model for the 500 m bridge with no significant gain in the accuracy of response estimations.}
\label{fig:runtime_comparison}
\end{figure}

The main objective of the simplified model is to improve the computational performance of simulations while having a minimal impact on the accuracy of the results. In Figure \ref{fig:runtime_comparison} the computational runtimes for the commercial vehicle simulation case are compared between two methods (the heavy vehicle yields a very similar plot as well). The figure elaborates that while the runtime increases linearly in the simplified model, it grows exponentially when using the conventional approach for longer bridges. For instance, using a single Intel Core i5 CPU, the entire VBI simulation process for the 500 m long bridge takes $1.8$ sec using the simplified model, while the same process takes nearly $2,250.0$ sec using the conventional method (more than $1,000x$ slower). This dramatic runtime difference is resulted by the inner iterations of the conventional approach (see Algorithm \ref{alg:conventional}) that guarantee the compatibility. Within this iteration, the entire bridge model has to be analyzed repeatedly for the modified interaction force as long as the stopping criterion is not met, which is computationally very expensive. This is a bottleneck for the numerical computation, especially when the bridge length increases or models with higher fidelity is of interest (i.e., MDF model grows in size). Alternatively, the simplified model fully decouples the bridge model from the vehicle systems, which yields a one-time bridge analysis (see Algorithm \ref{alg:simplified}). This significant speedup enables to perform VBI simulations for medium- to long-span bridges with fine spatial discretization, which is required for numerical studies on crowdsensing-based health monitoring. 

\section{Conclusions}

In this paper, a modified simulation algorithm was proposed for vehicle-bridge interaction (VBI) problems concerning medium- to long-span bridges with random traffic excitation. Our main contribution is the result that as the bridge flexibility increases (longer spans), the degree of coupling between the vehicle and the bridge reduces notably. Conventional VBI simulation algorithms require iterations within each time step in order to reach a desired level of compatibility between the vehicle and the bridge, which is computationally expensive. We show that the proposed simple, decoupled model is efficient for simulations of the vehicle-bridge interacting systems in such cases, with an accuracy that increases with bridge flexibility. In particular, the theoretical analysis showed that the response of a coupled continuous beam and vehicle setup subject to a random load becomes more independent to the vehicle dynamics as the bridge mass grows and the stiffness reduces. Therefore, for longer or flexible bridges, the dynamics are practically independent. Moreover, the numerical simulation validated that the bridge size and traffic load intensity both affect the accuracy of the bridge vibration estimations using the simplified model. For commercial vehicles, the simplified method yields accurate response estimations. In the case of a heavy vehicle with a natural frequency near the bridge's fundamental frequency, e.g., heavy vehicles and flexible bridges, the error associated with the simplified model is noticeable. In terms of the computational cost, a comparative study showed that the cost of the conventional model behaves exponentially while the cost of the simplified model is linear. 

\section{Acknowledgments}\label{S:Acknowledgments}

Research funding is partially provided by the National Science Foundation through Grant CMMI-1351537 by the Hazard Mitigation and Structural Engineering program, and by a grant from the Commonwealth of Pennsylvania, Department of Community and Economic Development, through the Pennsylvania Infrastructure Technology Alliance (PITA).

\bibliographystyle{unsrtnat} 
\bibliography{references}

\end{document}